\documentclass[aps,pre,reprint,superscriptaddress,footinbib,showpacs,floatfix]{revtex4-1}

\usepackage{comment}
\usepackage{graphicx}
\usepackage{latexsym}
\usepackage{xcolor}
\usepackage{amsmath}
\usepackage{hyperref}
\usepackage{multirow}
\usepackage{booktabs}
\usepackage{textcomp}
\usepackage{placeins} 
\usepackage{braket}
\usepackage{float}
\usepackage{ulem} 
\usepackage{appendix}
\usepackage[makeroom]{cancel}

\usepackage{bbm}
\usepackage{bm}
\usepackage{lipsum}

\usepackage{amsfonts,amssymb}

\newcommand{\Eqref}[1] {Eq.~\ref{#1}}
\renewcommand{\eqref}[1] {\ref{#1}}
\newcommand{\Figref}[1] {Fig.~\ref{#1}}
\newcommand{\figref}[1] {\ref{#1}}
\newcommand{\appref}[1] {Appendix~\ref{#1}}
\newcommand{\secref}[1] {Section~\ref{#1}}

\newcommand{\n}[0]{\nonumber \\}

\newcommand{\kt}{k_\mathrm{B}T}
\newcommand{\ho}{\mathrm{h.o.}}

\renewcommand{\vec}[1]{\mathbf{#1}}
\newcommand{\uvec}[1]{\widehat{\vec{#1}}}

\newcommand{\ident}[0]{\mathbbm{1}}
\newcommand{\tp}[0]{\intercal}

\renewcommand{\hm}[1]{\widehat{#1}}
\newcommand{\vecmap}[0]{\mathrm{vec}}
\newcommand{\hatmap}[0]{\mathrm{hat}}

\newcommand{\tr}[0]{\mathrm{tr}}
\newcommand{\cayeul}[1][]{\vec{f}_{\mathrm{eul}#1}}
\newcommand{\eulcay}[1][]{\vec{f}_{\mathrm{eul}#1}^{-1}}
\newcommand{\pmap}[0]{\mathcal{P}}

\newcommand{\totaltrans}[1]{\bar{#1}}

\newcommand{\ham}[0]{\mathcal{H}}

\newcommand{\dd}[0]{\mathrm{d}}

\newcommand{\cay}[1]{\mathrm{cay}\big(#1\big)}
\newcommand{\icay}[1]{\mathrm{cay}^{-1}\big(#1\big)}
\newcommand{\eul}[1]{\mathrm{eul}\big(#1\big)}
\newcommand{\ieul}[1]{\mathrm{eul}^{-1}\big(#1\big)}

\newcommand{\triad}[0]{\mathcal{T}}
\newcommand{\triadph}[0]{\mathcal{T}^{\phantom{\tp}}}
\newcommand{\triadtp}[0]{\mathcal{T}^{\tp}}
\newcommand{\rot}[0]{\mathcal{R}}
\newcommand{\drot}[0]{\mathcal{D}}
\newcommand{\srot}[0]{\mathcal{S}}
\newcommand{\rotph}[0]{\mathcal{R}^{\phantom{\tp}}}
\newcommand{\drotph}[0]{\mathcal{D}^{\phantom{\tp}}}
\newcommand{\srotph}[0]{\mathcal{S}^{\phantom{\tp}}}
\newcommand{\rottp}[0]{\mathcal{R}^\tp}

\newcommand{\srottp}[0]{\mathcal{S}^\tp}

\newcommand{\tse}[0]{\mathcal{\tau}}
\newcommand{\tseph}[0]{\mathcal{\tau}^{\phantom{\tp}}}
\newcommand{\tseinv}[0]{\mathcal{\tau}^{-1}}

\newcommand{\gse}[0]{g}
\newcommand{\gseph}[0]{\gse^{\phantom{-1}}}

\newcommand{\sse}[0]{s}

\newcommand{\sseinv}[0]{\sse^{-1}}

\newcommand{\dse}[0]{d}
\newcommand{\dseph}[0]{\dse^{\phantom{-1}}}

\newcommand{\gcg}[1]{g^{(#1)}}

\newcommand{\scg}[1]{s^{(#1)}}

\newcommand{\dcg}[1]{d^{(#1)}}

\newcommand{\vOm}[1][]{\vec{\Omega}_{#1}}
\newcommand{\vOms}[1][]{\vec{\Omega}_{0#1}}
\newcommand{\vOmd}[1][]{\vec{\Omega}_{\Delta#1}}

\newcommand{\vTh}[1][]{\vec{\Theta}_{#1}}
\newcommand{\vThs}[1][]{\vec{\Theta}_{0#1}}

\newcommand{\vPhs}[1][]{\vec{\Phi}_{0#1}}
\newcommand{\vPhd}[1][]{\vec{\Phi}_{\Delta#1}}

\newcommand{\hOm}[1][]{\hm{\Omega}_{#1}}
\newcommand{\hOms}[1][]{\hm{\Omega}_{0#1}}
\newcommand{\hOmd}[1][]{\hm{\Omega}_{\Delta#1}}

\newcommand{\hPhs}[1][]{\hm{\Phi}_{0#1}}
\newcommand{\hPhd}[1][]{\hm{\Phi}_{\Delta#1}}

\newcommand{\tOmd}[1][]{\bar{\vec{\Omega}}_{\Delta#1}}

\newcommand{\tThs}[1][]{\bar{\vec{\Theta}}_{0#1}}
\newcommand{\tThd}[1][]{\bar{\vec{\Theta}}_{\Delta#1}}

\newcommand{\vv}[1][]{\vec{v}_{#1}}
\newcommand{\vvs}[1][]{\vec{v}_{0#1}}
\newcommand{\vvd}[1][]{\vec{v}_{\Delta#1}}

\newcommand{\vs}[1][]{\vec{s}_{#1}}
\newcommand{\vd}[1][]{\vec{d}_{#1}}

\newcommand{\vX}[1][]{\vec{X}_{#1}}
\newcommand{\vXs}[1][]{\vec{X}_{0#1}}
\newcommand{\vXd}[1][]{\vec{X}_{\Delta#1}}

\newcommand{\vY}[1][]{\vec{Y}_{#1}}
\newcommand{\vYs}[1][]{\vec{Y}_{0#1}}
\newcommand{\vYd}[1][]{\vec{Y}_{\Delta#1}}

\newcommand{\tX}[0]{\bar{\vec{X}}}
\newcommand{\tXs}[0]{\bar{\vec{X}}_{0}}

\newcommand{\tY}[0]{\bar{\vec{Y}}}
\newcommand{\tYs}[0]{\bar{\vec{Y}}_{0}}
\newcommand{\tYd}[0]{\bar{\vec{Y}}_{\Delta}}

\newcommand{\tPsi}[1]{\bar{\vec{\Xi}}^{(#1)}}

\newcommand{\tPsid}[1]{\bar{\vec{\Xi}}_{\Delta}^{(#1)}}

\newcommand{\vYcgs}[2][]{\vec{Y}_{0#2}^{(#1)}}
\newcommand{\vYcgd}[2][]{\vec{Y}_{\Delta#2}^{(#1)}}

\newcommand{\tYcgs}[1]{\bar{\vec{Y}}_{0}^{(#1)}}
\newcommand{\tYcgd}[1]{\bar{\vec{Y}}_{\Delta}^{(#1)}}

\newcommand{\cgid}{q}

\newcommand{\kmaptot}[0] {\bar{A}^{(k)}}
\newcommand{\kmaptotinv}[0] {\left.\bar{A}^{(k)}\right.^{-1}}

\newcommand{\Ncg}[1]{N^{(#1)}}

\newcommand{\comp}[2]{(#1,#2)}
\newcommand{\accu}[2]{[#1,#2]}

\newcommand{\tw}{\mathrm{Tw}}

\newcommand{\ceff}{C_\mathrm{eff}}
\newcommand{\lb}{l_\mathrm{B}}
\newcommand{\lt}{l_\mathrm{T}}

\begin{document}

\title{Systematic Coarse-Graining of Sequence-Dependent Structure and Elasticity of Double-Stranded DNA}
\author{Enrico Skoruppa}
\email{esk.phys@gmail.com}
\affiliation{
    Cluster of Excellence Physics of Life, Technische Universit{\"a}t Dresden, Arnoldstraße 18, 01307 Dresden, Germany
}
\author{Helmut Schiessel}
\affiliation{
    Cluster of Excellence Physics of Life, Technische Universit{\"a}t Dresden, Arnoldstraße 18, 01307 Dresden, Germany
}
\affiliation{
    Institut f{\"u}r Theoretische Physik, Technische Universit{\"a}t Dresden, 01062 Dresden, Germany
}

\date{\today}


\begin{abstract}

Coarse-grained models have played an important role in the study of the behavior of DNA at length scales beyond a few hundred base pairs. Traditionally, these models have relied on structurally featureless and sequence-independent approaches, such as the twistable worm-like chain. However, research over the past decade has illuminated the substantial impact of DNA sequence even at the kilo-base pair scale. Several robust sequence-dependent models have emerged, capturing intricacies at the base pair-step level. Here we introduce an analytical framework for coarse-graining such models to the $2$ to $40$-base pair scale while preserving essential structural and dynamical features. These faithful coarse-grained parametrizations enable efficient sampling of large molecules. Rather than providing a fully parametrized model, we present the methodology and software necessary for mapping any base pair-step model to the desired level of coarse-graining. Finally, we provide application examples of our method, including estimates of the persistence length and effective torsional stiffness of DNA in a setup mimicking a freely orbiting tweezer, as well as simulations of intrinsically helical DNA.

\end{abstract}

\maketitle


\section{Introduction}
\label{sec:intro}

It is well established that the base sequence of DNA carries significance beyond its encoding into amino acids and the biochemical recognition of sequence motives via DNA-binding proteins. Both structure and elasticity of individual stretches of DNA have long since been shown to exhibit modulations vis-\`{a}-vis the underlying base pair sequence~\cite{olso98,lank00,lank03}. Such innate mechanical features constitute a rugged landscape of mechanical resistance that favors particular deformations at specific locations. For example, certain sequences display pronounced intrinsic curvature, i.e., the curvilinear center line is bent even in the absence of thermal fluctuations or externally induced deformations~\cite{Hagerman1986,Nelson1987,trif87,Koo1986,call88,Schellman1995,Peters2010}. Achieving a collectively curved contour over a stretch of DNA is easiest at places where the molecule is already naturally bent. At these places adhering to the innate direction of curvature is strongly favored. A prominent example is nucleosome positioning, i.e., the sequence preference for the wrapping of 147 base pairs around histone octamers in $1.7$ superhelical turns. Given the persistence length of roughly 150 bp the wrapping constitutes significant elastic deformation. Consequently, the consideration of structure and elasticity has been successful in the theoretical evaluation of preferred nucleosome-wrapped sequences~\cite{Morozov2009,EslamiMossallam2016,Neipel2020}.

At large length scales the elastic response of DNA is well described by homogeneous semi-flexible models and in particular the worm-like chain (WLC)~\cite{krat49,doi86,hage88,bust94,mark95} or the twistable worm-like chain (TWLC) when observables involving torsional and topological properties are considered~\cite{mark94,moro98,lipf10,brou18,nomi19,ott20,gao21,vand22,skor22}. However, even phenomena involving thousands of base pairs may exhibit appreciable sequence-specific behavior. For example, Kim et al.~\cite{kim18} demonstrated that DNA plectonemes tend to concentrate at specific sequence-encoded locations, which may carry profound implications for the spatial distribution of topological strain throughout the chromosome. Understanding the sequence-specific mechanical behavior of DNA may be crucial for comprehending the structural dynamics and functional fidelity of the chromosome.

Despite rapid advancements, experimental methods providing resolutions that reveal sequence-specific dynamics remain limited to date.
Accordingly, the utility of molecular simulations as effective computational microscopes remains uncontested. Atomistic simulations have been successful in replicating local mechanical properties of DNA~\cite{ivan15,zgar15,gali16}, and their application has been instrumental in unraveling a multitude of phenomena involving both bare DNA~\cite{lank06,maff10,mitc11,lieb15,vree19,Kim2021,Dohnalov2021,Yoo2021} and DNA-protein complexes~\cite{Saurabh2016,Liao2019,Yoo2020,Huertas2021,Coshic2024}. However, the computational cost associated with this level of detail is prohibitive for the study of systems involving more than a few hundred base pairs. 

Implicit solvent simulations have significantly extended the attainable length scale~\cite{Pyne2021}; however, reaching the length scales relevant to typical single-molecule experiments has thus far only been possible through the development of coarse-grained models. These models reduce the complexity of DNA to the necessary components for the particular phenomenon of interest, thereby reducing computational expense which in turn enhances scalability. A large variety of such models have emerged in recent years~\cite{knot07,ould10,sulc12,brac14,koro14,snod15,fosa16} some of which feature structural and elastic sequence-dependence~\cite{dans10,hink13,free14b,asse22}.

\begin{figure}[t]
\centering
\includegraphics[width=8.6cm]{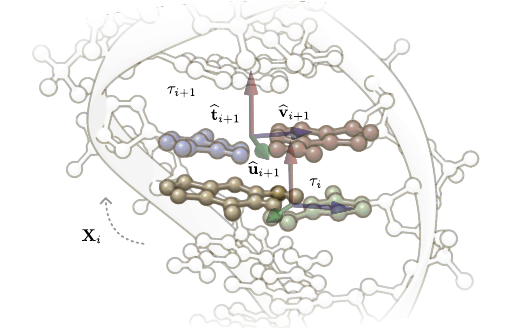}
\caption{Illustration of the rigid base pair description of DNA. Base pairs—depicted in blue-red and yellow-green—are captured by a single right-handed reference frame that captures the position and local orientation of the base pair. The relative orientations and positions of adjacent base pairs are parametrized in terms of six-vectors ($\vX[i]$) consisting of three rotational and three translational parameters.}
\label{fig:rbp} 
\end{figure}

A particularly prevalent way of coarse-graining DNA is the rigid base pair (RBP) model~\cite{lank09_rigidbp}, which represents each base pair as a single rigid body (see \Figref{fig:rbp}). Relative rotations and translations between consecutive base pairs are parametrized in terms of six degrees of freedom: three rotational (tilt, roll, and twist) and three translational (shift, slide, and rise). Incidentally, these degrees of freedom coincide with the most common way of classifying local elastic properties of DNA~\cite{hass95,olso01,lu03,lave09_curves,li19}. In corresponding models, the sequence-dependent intrinsic structure is included via the ground states of the six respective degrees of freedom. Deformations away from this ground state are commonly penalized with a quadratic elastic energy~\cite{lank03,olso98,petk14,sharma23}~\footnote{Some of these models are technically super-sets of RBP models, that include additional degrees of freedom. The corresponding RBP model is in these cases obtained by straight-forward marginalization of the excess degrees of freedom.}. This should be viewed as the lowest-order expansion of a generic underlying elastic energy. Truncation to quadratic order is warranted due to the stiff nature of double-stranded DNA. A given parametrization of an RBP model is fully characterized by a set of ground state coordinates and a stiffness matrix. 

To obtain sufficient statistics on systems involving thousands of base pairs the molecules under consideration are frequently coarse-grained to significantly larger sub-units than the basepair-step resolution of the RBP~\cite{volo97,lepa15,nomi17,ott20,skor22,vand22}. Not only do these coarser descriptions reduce the number of degrees of freedom, but coarse-graining also reduces the stiffness of the effective potentials, such that Molecular Dynamics simulations may be executed with larger time-steps~\cite{brac14,nomi17,ott20} and Markov Chain Monte Carlo simulations may feature larger cluster moves~\cite{klen91,liu08a,skor22}. 

Many studies rely on top-down parametrizations for coarse-grained descriptions, that utilize the homogenous TWLC model which has only two free parameters: the bending stiffness $A$
and the twist stiffness $C$.
These parameters are usually adopted from experimental measurements or higher resolution simulations, yielding a bending stiffness of about $40$-$55$~nm~\cite{bust94,stri00b,lipf11,mitc17,vand22} and a torsional stiffness in the range of $60$ to $110$~nm~\cite{fuji90,moro97,moro98,mosc09,lipf10,brya12,krie17b,gao21,vand22} under physiological ionic conditions. Alternatively, there have been several works providing schemes of analytically coarse-graining RBP parameters to TWLC parameters~\cite{beck07,skor21,guti23}. While these approaches yield parametrizations that give excellent agreement with the collective behavior of DNA molecules at large length scales, they are devoid of the structural and dynamic sequence features outlined before.

In this work, we present a systematic procedure for coarse-graining elastic RBP models to any resolution. The coarse-graining procedure entails a parameter transformation into a self-similar system, where the functional form of the model remains unchanged.

The article is structured as follows: The first half is dedicated to the description of the model and the development of the coarse-graining procedure. Section~\ref{sec:rbp} introduces the RBP model framework and casts the model in the form that is most convenient for the coarse-graining procedure. The coarse-graining scheme is detailed in Section~\ref{sec:cg_procedure}, and the specifics of the parameter transformation are provided in Section~\ref{sec:additive_fields}. A Python implementation for the parameter transformation is available at https://github.com/eskoruppa/PolyCG.

The second half of the paper presents various benchmark results assessing the effectiveness of the coarse-graining procedure. Section~\ref{sec:local_sampling} examines how well the coarse-grained system replicates individual distributions within unrestrained ensembles. Section~\ref{sec:lb} discusses the accuracy of reproducing length scale-dependent persistence lengths. Additionally, we simulate a freely orbiting magnetic tweezer setup for an experimentally studied sequence in Section~\ref{sec:fomt}, and analyze the behavior and response of helically curved DNA sequences in Section~\ref{sec:helix}. Finally, the paper concludes by summarizing the observations, reflecting on the potential impact of the work, and discussing its potential applications.


\section{Theory}
\label{sec:theory}


\subsection{Rigid Base Pair model}
\label{sec:rbp}

In the rigid base pair description of DNA, each base pair is treated as a rigid body associated with a couple $(\triad,\vec{r})$, capturing its orientation and position, respectively. Triads $\triad$ are right-handed reference frames that reflect the local geometry of the base pair in question, i.e., the orientation of the quasi-planar Watson-Crick base pairs, and the location of groves and backbones relative to the center of mass of the respective base pair. While exact definitions may vary across different conventions, most implementations—based on atomistic~\cite{lave09_curves,li19,voorspoels2023} and coarse-grained descriptions~\cite{freeman14,skoruppa17,chhabra20}—seek to generally encapsulate these geometric features. Throughout this work, we will define the three orthonormal basis vectors of each right-handed frame to populate the columns of the triad
\begin{equation}
    \label{eq:def_triads}
    \triad = \large[ \; \uvec{u} \; \uvec{v} \; \uvec{t} \; \large],
\end{equation}
such that global transformations act on triads in the same way as on any ordinary vector. 
Following standard convention~\cite{olso01}, $\uvec{t}$ is associated with the normal of the base pair plane (which for Watson-Crick base pairs is closely aligned with the curvilinear tangent of the molecule), $\uvec{u}$ roughly points from the base pair center of mass to the major groove, and $\uvec{v}$ is closely aligned with the vector connecting the two backbones. Note that, this definition classifies triads as elements of the rotation group, SO(3). 

Relative rotations and translations between adjacent base pairs  may be expressed in terms of six parameters---three rotational and three translational. When considering neighboring base pairs these components are commonly referred to as tilt, roll, and twist---for the rotational components---and shift, slide, and rise---for the translational components. To express these components independently of the global orientation of the molecule as a whole, relative rotations and translations are expressed in terms of the local material frame. The orientation of triad $\triad_{i+1}$ with respect to the frame of triads $\triad_i$ is given by
\begin{equation}
    \label{eq:accu_rot}
    \rot_{i} = \triadtp_i \triadph_{i+1}.
\end{equation}
These transformations between adjacent frames will occasionally be referred to as junctions or junction transformations. For this work, we will employ a definition of tilt, roll, and twist as the components of the rotation vector $\vec{\Omega}$ also known as Euler vector---associated with the rotation matrix $R$. The relationship
between the rotation matrices and rotation vectors is provided by the Euler map, which is explicitly given by Rodrigues' rotation formula (see \appref{app:rodrigues}) or equivalently by the exponential map
\begin{equation}
    \label{eq:expmap}
    \rot_i= \exp \hm{\Omega}_{i},
\end{equation}
where $\exp$ indicates the matrix exponential and the antisymmetric generators of rotation
\begin{equation}
    \label{eq:def_antisymmetric_tensor}
    \hOm 
    =
    \begin{pmatrix}
        0 & -\Omega_t & \Omega_v \\
        \Omega_t & 0 & -\Omega_u \\
        -\Omega_v & \Omega_u & 0 \\
    \end{pmatrix},
\end{equation}
are elements of the Lie algebra $\mathfrak{so}(3)$. The components $\Omega_u$, $\Omega_v$, $\Omega_t$—tilt, roll, and twist respectively—are simply the entries of the rotation vector 
$\vOm^\tp = \begin{pmatrix} \Omega_u & \Omega_v & \Omega_t\end{pmatrix}$. Hence, there is a one-to-one correspondence between rotation generators ($\mathfrak{so}(3)$) and rotation vectors ($\mathbb{R}^3$),
\begin{equation}
    \label{eq:hatmap}
    \hatmap \left(\vOm\right) = \hOm, 
    \quad \mathrm{and} \quad 
    \vecmap\left(\hOm\right) = \vOm,
\end{equation}
which we will refer to as hat-map and vector-map, respectively. 
An important relationship is that under the action of the vector map the Lie Bracket of $\mathfrak{so}(3)$ turns into the cross-product,
\begin{equation}
    \label{eq:liebracket}
    \vecmap ([\hm{\Omega}_1,\hm{\Omega}_2]) 
    = 
    \vec{\Omega}_1 \times \vec{\Omega}_2 
    = 
    \hm{\Omega}_1 \vec{\Omega}_2,
\end{equation}
where $\vOm[1]$ and $\vOm[2]$ are arbitrary rotation vectors.
A commonly employed alternative \cite{lank09_rigidbp,gonz13,thesis_davia,petk14,debruin18,sharma23} to the Euler-map is the Cayley map---also called the Euler-Rodrigues formula---which is closely related to the definition of unit quaternions. 
In \appref{app:cayley} we show how ground state and Gaussian elasticity, i.e., the stiffness matrix, can be transformed from the Cayley definition to the Euler definition and vice versa.   

Frequently, translations are expressed in the coordinate system of mid-step triads, which enables
a definition for translations independent of the choice of reference strand~\cite{olso01,lave09_curves,li19}. However, for the coarse-graining procedure outlined further below, it is advantageous to depart from this invariance and instead express relative translations within the frame of the first triad in each respective pair,
\begin{equation}
    \label{eq:def_transvec}
    \vec{v}_{i} = \triadtp_i \left( \vec{r}_{i+1} - \vec{r}_i \right).
\end{equation}
Despite the deviation from common convention, we will continue to refer to the translational components as shift, slide, and rise. Transformation of structure and elasticity from midstep- to triad-definition is discussed in \appref{app:midstep}.

Orientations and translations of triads may be cast in a unified representation within elements of the Special Euclidean Group~\cite{beck07}, SE(3), 
\begin{equation}
    \label{eq:def_tau}
    \tse_i 
    = 
    \begin{pmatrix}
        \triad_i & \vec{r}_i \\ 
        \vec{0}^\tp & 1 \\
    \end{pmatrix},
\end{equation}
where $\vec{0}$ is the three-dimensional null vector. Junction transformations then naturally contain the previously introduced rotation matrix and translation vector
\begin{align}
    \label{eq:def_g}
    \gse_i \equiv \tseinv_i \tseph_{i+1} 
    &=
    \begin{pmatrix}
        \triadtp_i\triadph_{i+1} & \triadtp_i ( \vec{r}_{i+1} - \vec{r}_i ) \\
        \vec{0}^\tp & 1 \\
    \end{pmatrix}
    \n
    &= 
    \begin{pmatrix}
        \exp \hOm[i] & \vv[i] \\ 
        \vec{0}^\tp & 1 \\
    \end{pmatrix},
\end{align}
and are therefore parametrized by six-component vectors 
\begin{equation}
    \label{eq:Xdef}
    \vX[i]^\tp 
    = 
    \begin{pmatrix}
        \vOm[i]^\tp & \vv[i]^\tp
    \end{pmatrix}.
\end{equation}
For convenience, we introduce a map between transformations $\gse_i$ and the corresponding parametrization vector
\begin{equation}
    \label{eq:def_pmap}
    \vX[i] \equiv \pmap \left(\gse_i\right). 
\end{equation}

The ground state of a particular molecule, which we also refer to as its structure, is characterized by certain sequence-specific junctions 
\begin{equation}
    \label{eq:def_si}
    \sse_i 
    \equiv 
    \begin{pmatrix}
        \srot_i & \vvs[,i] \\
        \vec{0}^\tp & 1 \\
    \end{pmatrix}
    =   
    \begin{pmatrix}
        \exp \hOms[,i] & \vvs[,i] \\ 
        \vec{0}^\tp & 1 \\
    \end{pmatrix}.
\end{equation}
Intrinsic bending and intrinsic twist are encoded in the static rotational parameters $\vOms$ while intrinsic translations—among which intrinsic rise is the most prominent—are contained in $\vvs$. Together these components are summarized in the ground state vector 
\begin{equation}
    \label{eq:vXs}
    \vXs[,i]^\tp 
    \equiv
    \pmap^\tp ( s_i )
    =
    \begin{pmatrix}
        \vOms[,i]^\tp & \vvs[,i]^\tp
    \end{pmatrix}.
\end{equation}

Static and dynamic components—fluctuations away from the ground state—are usually distinguished by splitting the junction six-vectors, \Eqref{eq:def_pmap}, into the respective components
\begin{equation}
    \label{eq:X_split_sd}
    \vX[i] = \vXs[,i] + \vXd[,i].
\end{equation}
However, for this work, it turns out to be a more prudent choice to split static and dynamic components at the transformation level rather than the vector level. This implies splitting the junctions $\gse_i$ into static components $\sse_i$ and a dynamic components $\dse_i$ as 
\begin{equation}
    \label{eq:gi_sd}
    \gse_i = \sse_i \dse_i,
\end{equation}
with
\begin{equation}
    \label{eq:def_di}
    \dse_i 
    \equiv  
    \begin{pmatrix}
        \drot_i & \vec{d}_i \\
        \vec{0}^\tp & 1 \\
    \end{pmatrix}
    =
    \begin{pmatrix}
        \exp \hPhd[,i] & \vd[i] \\ 
        \vec{0}^\tp & 1 \\
    \end{pmatrix}.
\end{equation}
To differentiate from the commonly used definition of fluctuations as the excess part of a single vector, we introduce a distinct notation for the components within our definition:
\begin{align}
    \vYs[,i] 
    &\equiv
    \pmap^\tp ( \sse_i )
    =
    \begin{pmatrix}
        \vPhs[,i]^\tp & \vs[i]^\tp
    \end{pmatrix},
    \\
    \vYd[,i] 
    &\equiv
    \pmap^\tp ( \dse_i )
    =
    \begin{pmatrix}
        \vPhd[,i]^\tp & \vd[i]^\tp
    \end{pmatrix}.
\end{align}
The static component is, of course, independent of the definition of fluctuations such that 
\begin{equation}
    \label{eq:equal_static}
    \vYs[,i] = \vXs[,i],
\end{equation}
i.e., $\vs[i] = \vvs[,i]$ and $\vPhs[,i] = \vOms[,i]$.

To fully capture the state of a given molecule containing say $N+1$ base pairs and $N$ junctions, we introduce the system-wide static and dynamic state-vectors
\begin{align}
    \label{eq:tXs_def}
    \tYs^\tp 
    &\equiv 
    \begin{pmatrix}
        \vYs[,0]^\tp & \vYs[,1]^\tp \ldots \vYs[,N-1]^\tp
    \end{pmatrix}
    \in \mathbb{R}^{6N},
    \\
    \label{eq:tXd_def}
    \tYd^\tp 
    &\equiv 
    \begin{pmatrix}
        \vYd[,0]^\tp & \vYd[,1]^\tp \ldots \vYd[,N-1]^\tp
    \end{pmatrix}
    \in \mathbb{R}^{6N}.
\end{align}

Fluctuations of the molecule away from its ground state, i.e., non-zero values of $\tYd$, are characterized by an elastic Hamiltonian $\ham(\tYd)$. Since double-stranded DNA is a rather stiff molecule and fluctuations at the base pair-step level are generally small, it is customary to consider the lowest non-trivial order expansion of $\ham(\tYd)$, which takes the shape of a quadratic form~\cite{olso98,lank03,lank09_rigidbp},
\begin{equation}
    \label{eq:X_energy_total}
    \beta \ham(\tYd) = \frac{1}{2} \tYd^\tp M_{\tY} \tYd,
\end{equation}
where, $M_{\tY}$, is a $6N\times6N$ stiffness matrix and the inverse temperature $\beta = 1 / \kt$ is assumed to be absorbed in the entries of the stiffness matrix. Unless the elasticity is assumed to be homogenous, this stiffness matrix will depend on the exact underlying sequence. Most studies definite these stiffness matrices for deformations in the $\vX$ definition~\cite{olso98,lank03,sharma23}. Transformations between these definitions and the $\vY$ definition considered here are discussed in \appref{app:algebra2group}.

Finally, we note that in many studies the elastic energy is assumed to be local, i.e., fluctuations between distinct junctions are fully decoupled~\cite{lank03,olso98}. In this case, the elastic energy simplifies to
\begin{equation}
    \label{eq:X_energy_local}
    \beta \ham(\tYd) = \frac{1}{2} \sum_{i=0}^{N-1} \vYd[,i]^\tp M_{\vY[i]} \vYd[,i],
\end{equation}
with $M_{\vY[i]}$ the individual $6\times6$ base pair-step stiffness matrices, which can usually be constructed from the $16$ canonical dimers (of which only 10 are truly unique~\cite{lank03,olso98}). Previous work, however, has shown, that the assumption of elastic locality is not satisfied within the framework of the rigid base pair description~\cite{lank09_rigidbp,noy12,skoruppa17,skor21,seger23,voorspoels2023,guti23} and that neighbor-couplings have to be considered to appropriately capture the long-range elastic properties~\cite{skor21,fosa21}. Whether these couplings imply physical interactions spanning across such lengths or whether they are mere artifacts introduced by the marginalization process from underlying higher resolution descriptions, is still debated~\cite{lank09_rigidbp}. Regardless of the origin of the couplings, they incontrovertibly have to be considered if the correct large-scale elastic behavior is sought to be captured. The coarse-graining procedure outlined in the next section naturally incorporates any such couplings (if present), since it is based on the most general (Gaussian) model, \Eqref{eq:X_energy_total}, which may in principle feature couplings at all length scales.


\subsection{Coarse-Graining Scheme}
\label{sec:cg_procedure}

\begin{figure}[t]
\centering
\includegraphics[width=8.6cm]{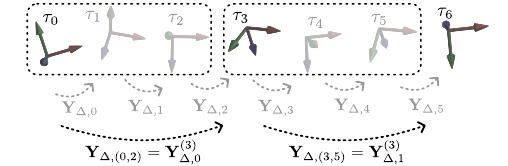}
\caption{Illustration of the coarse-graining scheme for $k=3$. The original chain of base pair reference frames is decimated such that only one in three triads is retained. Deformations of the coarse-grained system are expressed in terms of the junctions $\vYcgd[k]{,\cgid}$ connecting the remaining reference frames. These composite junctions each account for the accumulative fluctuation of three original junctions.}
\label{fig:cg_illustration} 
\end{figure}

We propose a $k$-step coarse-graining scheme that splits the chain into groups of $k$ triads and then eliminates all but the first triad in each group as is illustrated in~\Figref{fig:cg_illustration} for the case of $k=3$. The new—coarse-grained—junctions connect the remaining triads and therefore span over $k$ of the original junctions. This effectively corresponds to a $k$-fold increase in the discretization length. Assuming the original molecule to contain $N=\Ncg{k}k+1$ base pairs, the coarse-graining scheme reduces the system to the set of 
$\Ncg{k}+1$ triads $\{\tse_0, \tse_{k}, \cdots, \tse_{(\Ncg{k}-1)k}, \tse_{\Ncg{k}k} \}$
connected by the $k$-step junctions
\begin{equation}
    \label{eq:def_gcg}
    \gcg{k}_\cgid = \tseinv_{\cgid k} \tseph_{(\cgid+1)k},
\end{equation}
with $\cgid \in \{0\dots (\Ncg{k}-1)\}$. 
Following the same definition as in the base pair-step case, these junctions factor into static and dynamic components 
\begin{equation}
    \label{eq:composite_sd}
    \gcg{k}_\cgid = \scg{k}_\cgid \dcg{k}_\cgid,
\end{equation}
each of which is associated with coarse-grained static and dynamic parameters 
\begin{align}
    \label{eq:def_vYcgs}
    \vYcgs[k]{,\cgid} &\equiv \pmap \left( \scg{k}_\cgid \right), 
    \\
    \label{eq:def_vYcgd}
    \vYcgd[k]{,\cgid} &\equiv \pmap \left( \dcg{k}_\cgid \right).
\end{align}
Analogously, to the original system, the entire state of the reduced system is captured by the system-wide static and dynamics state-vectors $\tYcgs{k}$ and $\tYcgd{k}$ (see Eqs.~\eqref{eq:tXs_def} and \eqref{eq:tXd_def}). This coarse-graining scheme is reminiscent of the decimation mapping from real space renormalization group theory~\cite{kardar07}.
 
The coarse-graining procedure itself consists of identifying the coarse-grained ground state $\tYcgs{k}$ and the coarse-grained Hamiltonian $\ham(\tYcgd{k})$. Calculation of the former is straightforward and will be shown further below. The latter should be identified via the condition that the free energy of the system has to remain unchanged under the transformation, which is equivalent to requiring equal canonical partition functions
\begin{equation}
    \label{eq:equal_partition_function}
    Z 
    = \int \dd \tYd e^{-\beta \ham (\tYd) }
    = \int \dd \tYcgd{k} e^{-\beta \ham (\tYcgd{k})}.
\end{equation}
The main difficulty in the latter step stems from the requirement of finding a functional form for $\tYcgd{k}$ in terms of the original degrees of freedom $\tYd$.


\subsubsection{Composites}
Junctions between two arbitrary frames, as considered in \Eqref{eq:def_gcg}, may be written as a product of the intermediate junctions,
\begin{align}
    \label{eq:g_acc}
    g_{\accu{i}{j}} \equiv \prod_{l=i}^j g_l = \tseinv_{i} \tse_{j+1}. 
\end{align}
For such composites, we make the explicit distinction between the notation of composites which may be written as a product (for transformations) or sum (for vectors) of all corresponding intermediate junction elements, indicated by square bracket subscripts $\accu{i}{j}$, and those for which we do not a priori assume a trivial junction element decomposition to be possible, indicated by ordinary bracket subscripts $\comp{i}{j}$. For example, analogously to \Eqref{eq:def_g}, there will be a generator of rotation $\hOm[\comp{i}{j}]$ and a translation vector $\vv[\comp{i}{j}]$, such that
\begin{equation}
    \label{eq:def_gij}
    g_{\accu{i}{j}} =
    \begin{pmatrix}
        \exp \hOm[\comp{i}{j}] & \vv[\comp{i}{j}] \\ 
        \vec{0}^\tp & 1 \\
    \end{pmatrix}.
\end{equation}
While the matrix $g_{\accu{i}{j}}$ is by definition a product of single junction matrices, the composite vectors may generally not be written as the sum of the single-junction vectors ($\vOm[\comp{i}{j}] \neq \sum_{l=i}^j \vOm[l]$ and $\vv[\comp{i}{j}] \neq \sum_{l=i}^j \vv[l]$), i.e., they are nonadditive, except for certain elect cases (e.g., when the tangents, $\uvec{t}_i$, of all frames are aligned). 
For the construction of composites, it is most convenient to use the index range of the original junctions (from $i$ to $j$) rather than the index of the coarse-grained junction $q$. Bear in mind however that the relationship
\begin{equation}
    \gcg{k}_\cgid = g_{\accu{qk}{(q+1)k-1}}
\end{equation}
is implied.

\subsubsection{Coarse-Grained Ground State}
The coarse-grained ground state junctions may be written as a product of the contained original junctions (see \Eqref{eq:def_si}) which by employing the notation for accumulative transformations, \Eqref{eq:g_acc}, can be written as 
\begin{equation}
    \label{eq:sqk}
    \scg{k}_\cgid = s_{\accu{\cgid k}{(\cgid+1)k-1}} = \prod_{l=qk}^{(q+1)k-1} s_l.
\end{equation}
The corresponding ground state vector is then simply
\begin{equation}
    \label{eq:cggs}
    \vYcgs[k]{,\cgid} = \pmap \left( \scg{k}_\cgid \right).
\end{equation}


\subsubsection{Coarse-Grained Hamiltonian}
In general, the dynamic coarse-grained state vectors $\tYcgd{k}$ are not simply a linear combination of the components of the original dynamic state vectors $\tYd$. However, since fluctuations are generally small, we argue that expansion to linear order is warranted. The validity of this approximation will be further justified in the result section below. 

As a first step, we note that the state of the original system can be equivalently captured if we substitute, for each group of original junctions, a single junction by the coarse-grained junction $\vYcgd[k]{,\cgid}$ representing the respective group.

\begin{figure}[t]
\centering
\includegraphics[width=8.6cm]{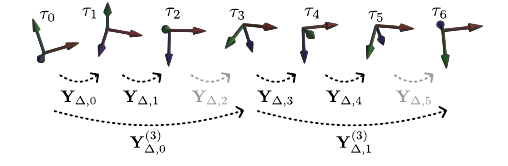}
\caption{
    This example configuration consisting of seven triads illustrates that the full state can be captured by a mixed set of original and coarse-grained dynamic junction parameters. In this particular example, a particular state of the system is fully parametrized by the set $\{ \vYd[,0], \vYd[,1], \vYcgd[k]{,0}, \vYd[,3], \vYd[,4], \vYcgd[k]{,1} \}$ (assuming that the ground state components are also given).
}
\label{fig:change_of_basis} 
\end{figure}
For example, a particular configuration of the system depicted in \Figref{fig:change_of_basis} can either be captured by the set of original deformations
$\{ \vYd[,0], \vYd[,1], \vYd[,2], \vYd[,3], \vYd[,4], \vYd[,5] \}$ or equivalently by the set $\{ \vYd[,0], \vYd[,1], \vYcgd[k]{,0}, \vYd[,3], \vYd[,4], \vYcgd[k]{,1} \}$, where the elements $\vYd[,2]$ and $\vYd[,5]$ are replaced by the coarse-grained junctions $\vYcgd[k]{,0}$ and $\vYcgd[k]{,1}$, respectively.

Arranging all the coarse-grained components to the right, this reformulated state vector takes the form 
\begin{equation}
    \label{eq:def_tPsid}
    \left.\tPsid{k}\right.^\tp
    = 
    \begin{pmatrix}
        \bar{\mathbf{Y}}_{\Delta, \mathrm{rem}}^\tp  & \left.\tYcgd{k}\right.^\tp
    \end{pmatrix},
\end{equation}
where $\bar{\mathbf{Y}}_{\Delta, \mathrm{rem}}$ contains all the remaining components, i.e., the ones that have not been substituted. 
In the example depicted in \Figref{fig:change_of_basis}, $\bar{\mathbf{Y}}_{\Delta, \mathrm{rem}}$ and $\tYcgd{k}$ are given by
\begin{align}
    \bar{\mathbf{Y}}_{\Delta, \mathrm{rem}}^\tp &= 
    \begin{pmatrix}
        \vYd[,0]^\tp & \vYd[,1]^\tp & \vYd[,3]^\tp & \vYd[,4]^\tp
    \end{pmatrix},
    \\
    \left.\tYcgd{k}\right.^\tp &= 
    \begin{pmatrix}
        \left.\vYcgd[k]{,0}\right.^\tp & \left.\vYcgd[k]{,1}\right.^\tp
    \end{pmatrix}.
\end{align}

Note that according to the aforementioned approximation the transformation from $\tYd$ to $\tPsid{k}$ is a change of basis and there will, thus, be a linear transformation $\kmaptot$ such that 
\begin{equation}
    \label{eq:tPsid_lintrans}
    \tPsid{k} = \kmaptot \tYd.
\end{equation}

With this transformation in hand we can transform the elastic energy, \Eqref{eq:X_energy_total}, as
\begin{align}
    Z 
    &= 
    \int \dd \tYd e^{-\frac{1}{2} \tYd^\tp M_{\tY} \tYd}
    \n
    &=
    \int \frac{\dd \tPsid{k}}{\det \kmaptot)} e^{-\frac{1}{2} \left(\kmaptotinv \tPsid{k} \right)^\tp M_{\tY} \left(\kmaptotinv \tPsid{k}\right)}
    \n
    &= 
    \int \dd \tPsid{k} e^{-\frac{1}{2} \left.\tPsid{k}\right.^\tp M_{\tPsi{k}} \tPsid{k} - \log\det \left(\kmaptot\right)},
\end{align}
where the the transformed stiffness matrix $M_{\tPsi{k}}$ is given by
\begin{equation}
    M_{\tPsi{k}} = \left(\kmaptotinv\right)^\tp M_{\tY} \kmaptotinv,
\end{equation}
and $(\det \kmaptot)^{-1}$ is the Jacobian of the transformation. 

All that is left to do to obtain the stiffness matrix $M^{(k)}$ of the coarse-grained system is to integrate out the remaining original degrees of freedom, $\bar{\mathbf{Y}}_{\Delta, \mathrm{rem}}$, which is equivalent to marginalizing the stiffness matrix. This can, for example, be achieved by taking the Schur complement of the matrix $M_{\tPsi{k}}$ with respect to $\tYcgd{k}$. By construction, the resulting elastic energy of the coarse-grained system will again be a quadratic form, i.e., it will have the same functional form as the original system.


\subsection{Composite Transformation}
\label{sec:additive_fields}

In this section, we derive the composite transformation $\bar{A}^{(k)}$ from \Eqref{eq:tPsid_lintrans}. The first step will be to express dynamic composite vectors $\vYd[,\comp{i}{j}]$ in terms of the original dynamic vectors $\{\vYd[,i], \ldots, \vYd[,j]\}$, and after proper expansion we identify the linear transformation for single composites
\begin{equation}
    \label{eq:vYdcomp_lintrans}
    \vYd[,{\comp{i}{j}}] = \sum_{l=i}^j A_l^{(i,j)} \vYd[,l].
\end{equation} 
Finally, these transformations are combined to construct the full system transformation $\bar{A}^{(k)}$.

\subsubsection{Dynamic Composite}
We seek to express the dynamic composites 
\begin{align}
    \label{eq:dij}
    d_{\comp{i}{j}} =
    \begin{pmatrix}
        \drot_{\comp{i}{j}} & \vd[\comp{i}{j}] \\ 
        \vec{0}^\tp & 1 
    \end{pmatrix},
\end{align}
with $\drot_{\comp{i}{j}} = \exp \hPhd[,\comp{i}{j}]$, in terms of the single junction static and dynamic components
\begin{equation}
    \label{eq:def_si_di_repeat}
    \sse_i 
    = 
    \begin{pmatrix}
        \srot_i & \vs[i] \\
        \vec{0}^\tp & 1 \\
    \end{pmatrix},
    \quad
    \dse_i 
    = 
    \begin{pmatrix}
        \drot_i & \vd[i] \\
        \vec{0}^\tp & 1 \\
    \end{pmatrix},
\end{equation}
where $\srot_i = \exp \hPhs[,i]$, and $\drot_i = \exp \hPhd[,i]$ (see Eqs.~\eqref{eq:def_si}, \eqref{eq:equal_static}, and \eqref{eq:def_di}).
Using \Eqref{eq:composite_sd} we can write $\dseph_{\comp{i}{j}} = \sseinv_{\accu{i}{j}} \gseph_{\accu{i}{j}}$ and after working out 
the right-hand side explicitly in terms of the components of Eqs.~\eqref{eq:def_si_di_repeat} one finds
\begin{equation}
    \label{eq:Dij}
    \drot_{\comp{i}{j}} = \srottp_{\accu{i}{j}} \rotph_{\accu{i}{j}},
\end{equation}
and 
\begin{align}
    \label{eq:dij}
    \vd[\comp{i}{j}] 
    &= \srottp_{\accu{i}{j}} \vv[{\comp{i}{j}}] - \srottp_{\accu{i}{j}} \vs[_{\comp{i}{j}}]
    \n
    &= \srottp_{\accu{i}{j}} \sum_{l=i}^j \left( \prod_{m=i}^{l-1} \left(\srot_m \drot_m \right) \left(\srot_l \vd[l] + \vs[l] \right) \right) 
    \n
    &\phantom{=} \qquad \qquad - \sum_{l=i}^{j} \srottp_{\accu{l}{j}} \vs[l].
\end{align}
We will calculate the rotational and translational components of $\vYd[,{\comp{i}{j}}]$ separately, starting with the former.


\subsubsection{Rotation}
\label{sec:ad_rot}

A straightforward calculation shows that \Eqref{eq:Dij} can be rewritten as
\begin{align}
    \drot_{\comp{i}{j}} 
    &= 
    \prod_{l=i}^{j} \left( \srot_{\accu{l+1}{j}}^\tp \drotph_l \srotph_{\accu{l+1}{j}} \right)
    \n
    &=
    \prod_{l=i}^{j} \exp \left( \srottp_{\accu{l+1}{j}} \hPhd[,l]^{\phantom{\tp}} \srotph_{\accu{l+1}{j}}\right),
\end{align}
where the last equality follows directly from the properties of matrix exponentials. 
To arrive at an expression for a single rotation matrix comprising all the fluctuating components, we make the approximation
\begin{equation}
    \label{eq:Dij_linear_approx}
     \drot_{(i,j)} 
     \approx \exp \left( \sum_{l=i}^{j} \srottp_{\accu{l+1}{j}} \hPhd[,l]^{\phantom{\tp}} \srot_{\accu{l+1}{j}} \right),
\end{equation}
which is equivalent to discarding all terms of higher than linear order in the respective series expansions (or equivalently, ignoring all commutators in the Baker-Campbell-Hausdorff formula). Such an expansion is justified, because we separated the fluctuating components, which are assumed to be small, from the static components that assume appreciable values—especially intrinsic twist. This explains the unconventional choice of expressing the rigid base pair model in $\tY$ coordinates rather than the usual $\tX$ coordinates. The fluctuational component of the composite step rotation matrix can therefore be written as a sum of transformed—approximately additive—generators
\begin{align}
    \hm{\Phi}_{\Delta,l}^{'} 
    &= \srottp_{\accu{l+1}{j}} \hPhd[,l]^{\phantom{\tp}} \srotph_{\accu{l+1}{j}} \n
    &= \hatmap \left( \srottp_{\accu{l+1}{j}} \vPhd[,l]^{\phantom{\tp}}  \right),
\end{align}
where the second equality follows from general properties of rotation generators. Invoking the vector map, \Eqref{eq:hatmap}, shows that these transformed and approximately additive rotation vectors are obtained by rotating the original rotation vectors
\begin{equation}
   \vPhd[,l]^{'} = \srottp_{\accu{l+1}{j}} \vPhd[,l]^{\phantom{\tp}}.
\end{equation}
Finally, summing over all the terms is \Eqref{eq:Dij_linear_approx} yields the sought transformation for the rotational components
\begin{equation}
    \label{eq:expr_phaccu}
    \vPhd[,\comp{i}{j}] \approx \sum_{l=i}^j \srottp_{\accu{l+1}{j}} \vPhd[,l]^{\phantom{\tp}}.
\end{equation}


\subsubsection{Translation}
\label{sec:ad_trans}

To arrive at the expression for composite translations we expand all occurrences of the rotation matrices $\drot$ in \Eqref{eq:dij} to linear order
\begin{equation}
    \drot = \exp \hPhd \approx \ident + \hPhd,
\end{equation}
which is again warranted as we assumed the components of $\vPhd[,k]$ to be small. Discarding all terms of higher than linear order in any of the fluctuating components (both rotation and translation), one eventually arrives at (see \appref{app:comp_trans} for details),
\begin{align}
    \label{eq:comp_trans}
    \vd[\comp{i}{j}] 
    &\approx
    \sum_{l=i}^{j-1} \left[ \sum_{m=l+1}^j \hatmap\left( \srottp_{\accu{m}{j}}\vs[m]^\tp \right) \srottp_{\accu{l+1}{j}} \right] \vPhd[,l]
    \n
    &\phantom{=} + \sum_{l=i}^{j} \srottp_{\accu{l+1}{j}} \vd[l].
\end{align}


\subsubsection{Constructing the linear transformation}
\label{sec:transformation_scheme}

Jointly, the results of Eqs.~\eqref{eq:expr_phaccu} and \eqref{eq:comp_trans} may be summarized in the form of the linear transformation \Eqref{eq:vYdcomp_lintrans},
with
\begin{equation}
    \label{eq:Al_ij}
    A_l^{(i,j)} = 
    \begin{pmatrix}
        A_{l,\mathrm{rr}}^{(i,j)} & 0 \\
        A_{l,\mathrm{rt}}^{(i,j)} & A_{l,\mathrm{tt}}^{(i,j)}
    \end{pmatrix}
\end{equation}
and entries
\begin{align}
    A_{l,\mathrm{rr}}^{(i,j)} 
    &= \srottp_{\accu{l+1}{j}}
    \\
    A_{l,\mathrm{tt}}^{(i,j)} 
    &= \srottp_{\accu{l+1}{j}}
    \\
    A_{l,\mathrm{rt}}^{(i,j)} 
    &= \sum_{m=l+1}^j \hatmap\left( \srottp_{\accu{m}{j}}\vs[m]^\tp \right) \srottp_{\accu{l+1}{j}}.
\end{align}
To determine the transformation $\bar{A}^{(k)}$ from \Eqref{eq:tPsid_lintrans} we will first construct the matrix of basis change that transforms dynamic state vectors spanning over a single compound, i.e., from junction $i$ to junction $j$, from the original basis
\begin{equation}
    \tilde{\vec{Y}}_{\Delta, \comp{i}{j}}^\tp = 
    \begin{pmatrix}
        \vYd[,i]^\tp & \dots & \vYd[,j-1]^\tp & \vYd[,j]^\tp
    \end{pmatrix},
\end{equation}
to the basis in which the last entry is substituted by the respective dynamic compound vector
\begin{equation}
    \tilde{\vec{\Xi}}_{\Delta, \comp{i}{j}}^\tp = 
    \begin{pmatrix}
       \vYd[,i]^\tp & \dots & \vYd[,j-1]^\tp & \vYd[,\comp{i}{j}]^\tp
    \end{pmatrix}.
\end{equation}
This transformation takes the form
\begin{equation}
    \tilde{\vec{\Xi}}_{\Delta, \comp{i}{j}} = A^{(i,j)} \tilde{\vec{Y}}_{\Delta, \comp{i}{j}}
\end{equation}
with
\begin{equation}
    A^{(i,j)} = 
    \begin{pmatrix}
        \ident & 0 & \hdots & 0 & 0 \\
        0 & \ident & \hdots & 0 & 0 \\
        \vdots & \vdots & \ddots & \vdots & \vdots \\
        0 & 0 & \vdots & \ident & 0 \\
        A^{(i,j)}_i & A^{(i,j)}_{i+1} & \hdots & A^{(i,j)}_{j-1} & \ident
    \end{pmatrix}.
\end{equation}
The right-most entry in the bottom row is identity since $A^{(i,j)}_{j} = \ident$. The transformation of single compounds can be extended to the entire system via the block-diagonal matrix
\begin{equation}
    \bar{A}^{'(k)} = 
    \begin{pmatrix}
        A^{(0,k-1)} & 0 & \hdots & 0 \\
        0 & A^{(k,2k-1)} & \hdots & 0 \\
        \vdots & \vdots & \ddots & \vdots \\
        0 & 0 & \hdots & A^{([N^{(k)}-1]k,N^{(k)}k-1)}        
    \end{pmatrix}.
\end{equation}
Finally the $\bar{A}^{(k)}$ from \Eqref{eq:tPsid_lintrans} is obtained by rearranging the elements in $\bar{A}^{'(k)}$ via a permutation that arranges all the coarse-grained components to the back of the vector. This can be achieved by invoking a permutation matrix $P$\begin{equation}
     \bar{A}^{(k)} = P  \bar{A}^{'(k)}.
\end{equation}
Since $\bar{A}^{'(k)}$ is a block diagonal matrix with component blocks $A^{(i,j)}$, each of which satisfying $\det A^{(i,j)} = 1$, the transformation matrix $\bar{A}^{(k)}$ will also have unit determinant as per the orthogonality of permutation matrices.


\section{Results and Discussion}
\label{sec:results}

In this section, we assess the fidelity of the coarse-graining procedure by comparing various parameters sampled at different levels of coarse-graining. Our analysis relies on sequence-dependent ground state and stiffness parameters obtained from the cgNA+ model~\cite{sharma23,thesis_rahul}. This model extends the RBP model by incorporating additional degrees of freedom beyond the original six parameters (tilt, roll, twist, shift, slide, and rise). Specifically, it treats bases and phosphate groups as rigid bodies, resulting in additional six intra-base pair parameters (buckle, propeller, opening, shear, stretch, and stagger), along with six degrees of freedom encoding the orientation and position of the phosphate group relative to the corresponding base in each nucleotide. Stiffness matrices for the RBP model are obtained by marginalizing these excess degrees of freedom, employing appropriate Schur complements as outlined in previous work~\cite{lank09_rigidbp}.

In the cgNA+ model, rotations are represented using the Cayley map. However, our coarse-graining procedure necessitates parameters expressed in terms of the Euler map. We therefore transform the ground state and the stiffness matrices to this representation (further details provided in \appref{app:cayley}). Our approach assumes translations between pairs of triads to be defined in the frame of the first triad of the respective pair, while cgNA+ expresses translations in the corresponding midstep frame. Additionally, as discussed in \secref{sec:rbp}, we segregate static and dynamic components of rotations and translations at the transformation matrix level, rather than at the corresponding vector level, as commonly done in RBP models. To use the language and symbols introduced in this work we require the system to be parametrized in terms of $\tY$ coordinates instead of the more common $\tX$ coordinates. This transformation of the ground state and stiffness from midstep triad to triad definition of translations, along with the redefinition of fluctuating components, is achieved via the transformations derived in Appendices \ref{app:algebra2group} and \ref{app:midstep}. The outcome is a set comprising the ground state and a stiffness matrix ${(\tYs,M_{\tY})}$, which can then be transformed into coarse-grained sets ${(\tYcgs{k},M_{\tY }^{(k)})}$ using the scheme outlined in \secref{sec:transformation_scheme}.


\subsection{Comparison with unrestrained Monte Carlo Sampling}
\label{sec:local_sampling}

\begin{figure*}[t]
\centering
\includegraphics[width=17.6cm]{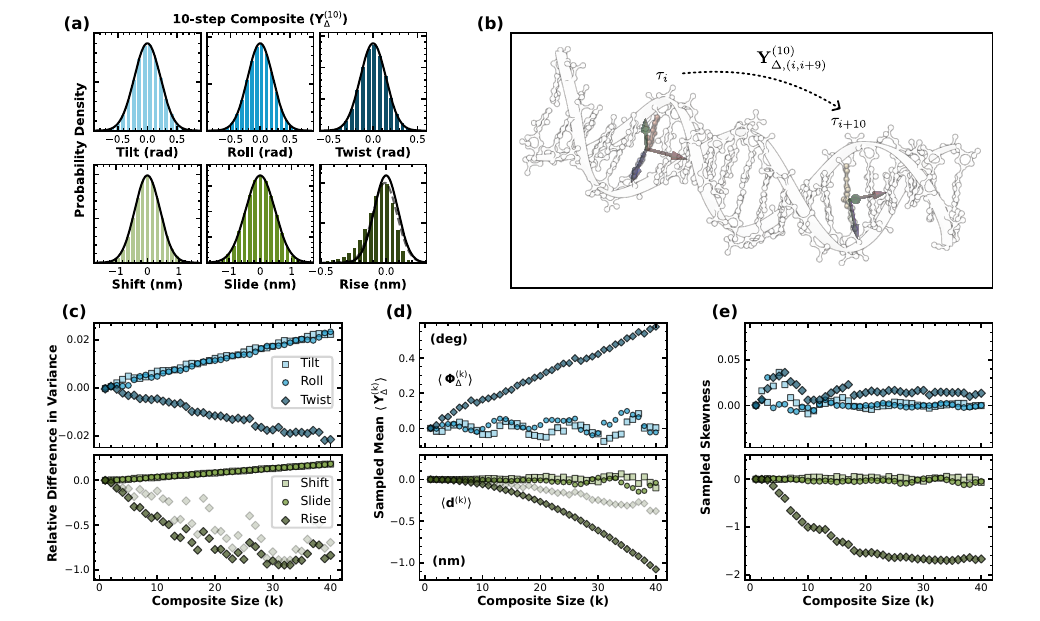}
\caption{Comparison between sampled and distributions resulting from the analytical coarse-graining procedure. (a) Histograms for the six degrees of freedom of a single $10$-step composite. The black lines are normal distributions with a variance taken from the stiffness matrix of the coarse-grained system. (b) Illustration of the coarse-grained system represented in panel (a). (c) Relative difference between the variance resulting from the coarse-graining procedure and the sampled variance according to \Eqref{eq:var_reldiff} for values of $k$ ranging from $2$ to $40$. Mean of skewness (see \Eqref{eq:skewness}) of the same distributions are displayed in panels (d) and (e), respectively. Transparent scatters in (c) and (d) correspond to Gaussian fits centered at the value of maximum likelihood.} 
\label{fig:local_composites} 
\end{figure*}
We first consider ensembles of unrestrained configurations ranging over single composite steps for composite sizes $k$ ranging from $2$ to $40$. For each value of $k$, we generated a sample of $2.5\times10^7$ unrestrained configurations drawn according to the canonical measure. Specifically, we drew ${\tYd \in \mathbb{R}^{6k}}$ from the multivariate Gaussian distribution
\begin{equation}
    \rho(\tYd) = \left(\frac{\det M_{\tY})}{(2\pi)^{6k}}\right)^{\frac{1}{2}} e^{-\frac{1}{2} \tYd^\tp M_{\tY} \tYd},
\end{equation}
where the inverse temperature ${\beta=(k_\mathrm{B} T)^{-1}}$ is absorbed in $M_{\tY}$. For each individual set, we then constructed a configuration consisting of $k+1$ triads ${\{\tau_0,\ldots,\tau_{k}\}}$. Using the known composite ground state transformation $s^{(k)}$ (see \Eqref{eq:sqk}) we calculated the dynamic composite components $\vYcgd[k]{} = \pmap(d^{(k)})$ as the parameter associated with the transformation 
\begin{equation}
    d^{(k)} = \left.s^{(k)}\right.^{-1} g^{(k)} = \left.s^{(k)}\right.^{-1} \tau_0^{-1} \tau_k.
\end{equation}
According to the assumption of approximate linearity of the composite transformation, the six individual components $\vYcgd[k]{}$—for the sake of simplicity we will also call them tilt, roll, twist, shift, slide, and rise—should be multivariate Gaussian distributed. Histograms of the sampled values for composites composed of $10$ junctions (${k=10}$) are shown in \Figref{fig:local_composites}(a). For comparison, we co-plot the distributions resulting from our analytically coarse-graining stiffness matrices ${M_{\tY }^{(k)} \in \mathbb{R}^6}$. The variances of individual degrees of freedom are obtained via marginalization. Following the construction of the $\vYcgd[k]{}$ as deviations away from the ground state, their mean is assumed to be zero. Both the assumption of Gaussianity and the agreement between sampled and predicted distribution are excellent in most components, except for the rise component, which exhibits pronounced left-skewness. We attribute this broken symmetry to the influence of bending fluctuation on the accumulative rise. On the one hand, an increased composite rise can only result from overstretching of the translational degrees of freedom. Reduction in composite rise, on the other hand, may result from a combination of contracted rise translations and local bending fluctuations analogous to the entropic spring behavior of flexible and semi-flexible polymers~\cite{doi86,mark95}.

For quantitative assessment we evaluated the variance, mean, and skewness of the six components of $\vYcgd[k]{}$ for all considered values of $k$; shown in Figures~\figref{fig:local_composites}(c), (d), and (e), respectively. To put differences in variance into perspective we show relative differences between sampled and analytically coarse-grained variances as given by 
\begin{equation}
    \label{eq:var_reldiff}
    \Delta \mathrm{var}_\mathrm{rel} = \frac{\mathrm{var}_\mathrm{cg} - \mathrm{var}_\mathrm{sampled}}{\mathrm{var}_\mathrm{sampled}}.
\end{equation}
Since $\langle \vYcgd[k]{} \rangle = 0$ by construction of the coarse-graining procedure, only the sampled mean values are displayed. Likewise, the analytical coarse-graining is constructed as a linear transformation which transforms the original Gaussian system into another Gaussian system. Therefore, the third central moment of the transformed system is assumed to be zero. Omitted higher-order contributions may lead to non-vanishing third central moments. \Figref{fig:local_composites}(e) shows the normalized sampled third central moments   
\begin{equation}
    \label{eq:skewness}
    \tilde{\mu}_3 = \frac{\left\langle \left(x - \langle x \rangle\right)^3 \right\rangle}{\left\langle \left( x - \langle x \rangle \right)^2 \right\rangle^{3/2}},
\end{equation}
which is sometimes referred to as Fisher's moment coefficient of skewness.

Rotational degrees of freedom display excellent agreement between sampled and predicted values, with relative differences in the variance of less than $2.5\%$ for the largest considered coarse-graining size ($k=40$). Mean and skewness of these values is generally small, with only the mean of the twist exhibiting increasing deviations, reaching about $0.58$~deg for the largest considered coarse-graining size. This value constitutes a mere $3\%$ of the standard deviation of the twist fluctuations at this level of coarse-graining. 

Deviations between transformed and sampled distributions are significantly larger for the translational degrees of freedom. While the relative difference in variance reaches about $20\%$ for the two lateral components shift and slide, deviations in rise compound to almost $100\%$ for the largest considered composite size. This large deviation stems from the aforementioned symmetry breaking in rise fluctuations that manifests in appreciable left-skewness of the distribution. As a direct consequence, one observes a lower mean and a significantly larger variance. The assumption of Gaussianity of compound-step rise is evidently not satisfied. Consequently, a higher-order energetic description would be necessary to accurately capture the behavior of rise fluctuations. While possible in principle, the development of such a model is beyond the scope of the present work. For closer comparison to the quadratic model, we fit the sampled rise distribution with a Gaussian centered around the most likely rise value (results shown as transparent scatters). This way of comparison reveals discrepancies to be less severe, especially for composite-step sizes smaller than about $k=10$.


\subsection{Persistence Lengths}
\label{sec:lb}

The bending and torsional persistence lengths are the canonical measures of DNA elasticity quantifying the mechanical response of the molecule at the mesoscale. There is a variety of different definitions for the bending persistence length~\cite{mitc17}. Here we are employing a definition based on the exponential decay of the tangent-tangent correlation function~\cite{mazur2007}
\begin{equation}
    \label{eq:tancor}
    \left\langle \uvec{t}_{n} \cdot \uvec{t}_{n+m} \right\rangle = e^{-\frac{am}{\lb}},
\end{equation}
where $m$ is the curvilinear distance expressed in base pair-steps, and $a$ is the discretization length, which we set to $a=0.34$~nm. The expectation brackets in \Eqref{eq:tancor} indicate simultaneous thermal- and sequence averages, i.e., the expression is averaged over all possible reference indices $n$, provided that the molecule contains at least $n+m$ base pairs. 

Regular exponential decay of the tangent-tangent correlation function, as indicated in \Eqref{eq:tancor}, is behavior exhibited only by semi-flexible polymers characterized by purely local elasticity~\cite{skor21}—which is, for example, the case for an elastic Hamiltonian of the form given by \Eqref{eq:X_energy_local}—and no structural features~\cite{noy12}. Previous work has shown that non-locality in the elastic energy, i.e., couplings between neighboring junctions and beyond, give rise to distinct length-scale dependence of the elastic properties~\cite{noy12,wu15,skor21,seger23,laer24}. Moreover, intrinsic bending components are known to give rise to additional deviations from the exponential behavior~\cite{noy12}. We, therefore, employ a length scale-dependent definition of the persistence length based on \Eqref{eq:tancor}, but evaluated for every $m$ individually~\cite{skor17} 
\begin{equation}
    \label{eq:lb}
    \lb(m) = \frac{-am}{\log\left\langle \uvec{t}_{n} \cdot \uvec{t}_{n+m} \right\rangle}.
\end{equation}

Analogous to \Eqref{eq:tancor} the twist-elasticity is characterized by the twist-correlation function
\begin{equation}
    \label{eq:twcor}
    \left\langle \sum_{i=n}^{n+m-1} \left[\vPhd[,i]\right]_3 \right\rangle = \left\langle \cos \left[\vPhd[,i]\right]_3 \right\rangle^{m} = e^{-\frac{am}{\lt}},
\end{equation}
where $\left[\vPhd[,i]\right]_3$ is the third component of the excess rotation $\vPhd[,i]$. The associated decay length, the torsional persistence length $\lt$, is related to the torsional stiffness by a factor of two: $\lt = 2C$~\cite{brac14}. Equation~\eqref{eq:twcor} again only holds for twist-storing polymers with purely local elastic couplings. Non-locality of the elastic energy breaks the monotonous decay of the twist-correlation function and introduces length-scale dependence, warranting a local definition analogous to the expression for the bending persistence length~\cite{skor17}
\begin{equation}
    \label{eq:twistcor}
    \lt(m) = \frac{-am}{\log\left\langle \sum_{i=n}^{n+m-1} \left[\Omega_{\Delta,i}\right]_3 \right\rangle}. 
\end{equation}

\begin{figure*}[ht!]
\centering
\includegraphics[width=17.6cm]{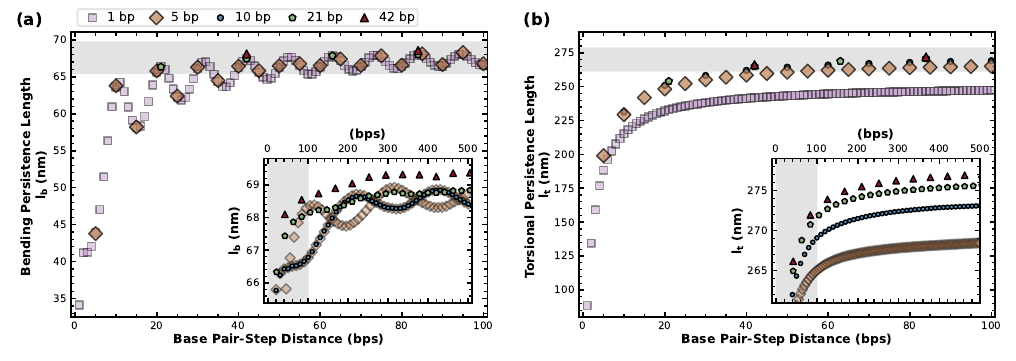}
\caption{Length scale-dependent bending (a) and torsional (b) persistence lengths of $7922$ base pair sequence. The considered molecule was sampled at varying coarse-grained resolutions as indicated by the different scatter symbols. Extended base pair-step distances within a narrowed range of persistence length are displayed in the insets. Gray-shaded regions indicate the ranges considered in the other figure.
}
\label{fig:lb} 
\end{figure*}

To highlight the utility of the coarse-graining procedure, we compare the original base pair resolution model to coarse-grained parametrizations of various resolutions. In particular, we considered composites spanning $5$, $10$, $21$, and $42$ base pairs, corresponding to roughly half, one, two, and four helical repeats, respectively. Rather than considering a single composite step as in the previous section, we simulated a $7922$~base pair sequence that has been employed in various experimental single-molecule studies~\cite{lipf10,krie18,vand22}. Following the procedure introduced in the previous section, we generated $10^8$ independent configurations for each resolution (i.e., for each value of $k$). 

Length scale-dependent bending persistence lengths for base pair-step distances ($m$) ranging from $1$ to $500$ are displayed in \Figref{fig:lb}. Consistent with previously reported findings, the simulation at base pair resolution ($m=1$), exhibits low persistence, i.e., enhanced flexibility, at short distances while asymptotically converging towards larger stiffness for large $m$~\cite{noy12,wu15,skor17,skor21,sege21,guti23,laer24}. Moreover, the behavior of $\lb(m)$ is further modulated by a sinusoidal oscillation of wavelength corresponding to the helical repeat length (approximately $10.5$~bp), which is a result of the base pair planes being slightly tilted on average relative to the helical axis (see for example Ref.~\cite{noy12}). Asymptotically, $\lb(m)$ converges to about $69$~nm. While this value is significantly larger than the literature values of $40$-$55$~nm, it is in line with previous studies reporting on the persistence length of cgNA+~\cite{thesis_rahul,laer24} and similar to values found with atomistic simulations~\cite{skor21} conducted with the parmbsc1 forcefield~\cite{ivan15} based on which cgNA+ is parametrized. 

In this work, our focus lies not on the accuracy of the underlying RBP parametrization, but rather on the fidelity of reproducing configurational fluctuations at the coarse-grained level. As depicted in \Figref{fig:lb}(a), the coarse-grained parametrizations yield persistence length values that closely align with those of the original single base pair resolution model.

The twist-persistence length values exhibit less agreement between simulations at different resolutions: at base pair resolution $\lt(m)$ converges to approximately $250$~nm ($C\approx125$~nm), whereas at lower resolutions $\lt(m)$ converges to somewhat larger values ($C\approx135-140$~nm). We propose that these discrepancies do not arise from intrinsic deficiencies in the coarse-graining procedure, but rather from an inaccurate definition of twist as the third component of the junction rotation vector. For twist to be additive, it should quantify rotation around the helical axis of the molecule. However, due to fluctuations in the tangents of the reference frame away from the molecular center line and helical variations in base planes, twist fluctuations at larger length scales are influenced by geometric and topological features~\cite{full78}. This point will be further illustrated in the following Section.


\subsection{Force extension and effective torsional stiffness}
\label{sec:fomt}

As an example for simulations closer related to experiments, we consider the setup of freely orbiting magnetic tweezers (FOMT)~\cite{lipf11} (see \Figref{fig:fomt}(a)). In such experiments, a single double-stranded DNA molecule is tethered between a flow cell surface and a superparamagnetic magnetic bead. Linear stretching forces $\vec{f}$ can be applied to the DNA tether by proxy of exposing the bead to magnetic field gradients induced by a cylindrically shaped magnet located above the bead. 

First, we will summarize the theory essential for rationalizing the experimental readout and extracting bending and torsional properties. Following this, we present the results obtained from extensive Monte Carlo simulations sampled at various resolutions.

\subsubsection{Force Extension}
Measurement of the hovering height of the bead relative to the surface gives access to the instantaneous tether extension $z$. Elongation of the molecule is opposed by an accompanying loss in entropy, which results in a characteristic force-dependent equilibrium. For sufficiently large forces ($f \gtrsim 0.3$~pN) small fluctuation theory provides a simple expression for the mean extension $\langle z\rangle$ in terms of force $f$, tether length $L$, and the bending persistence length $\lb$~\cite{mark94,mark95}
\begin{equation}
    \label{eq:forceext_highforce}
    \frac{\left\langle z \right\rangle}{L} = 1 - \frac{1}{2}\sqrt{\frac{\kt}{\lb f}} + \ho .
\end{equation}
To cover the full range of forces, several interpolation formulas have been brought forward that continuously connect the response at large forces to the behavior of ideal flexible polymers at low forces~\cite{mark95,bouc99}. The first and possibly simplest of these was suggested by Marko and Siggia~\cite{mark95}
\begin{equation}
    \label{eq:forceext_interp}
    \frac{\lb f}{\kt} = \frac{\langle z\rangle}{L}+\frac{1}{4\left(1-\langle z\rangle/L\right)^2}-\frac{1}{4}.
\end{equation}
These expressions give access to the bending persistence length by fitting measured  mean extensions for a range of forces. 

\subsubsection{Effective Torsional Stiffness}
Suitable positioning of the magnet allows for the monitoring of rotations of the bead around the force director field (for details see Ref.~\cite{lipf11}). Effectively, this gives access to the torsional fluctuations of the molecule as a whole, which can be applied to assess the molecule's innate torsional properties. However, rotation of the bead, or equivalently the molecular terminus, does not translate into an equal amount of accumulative twist strain distributed over the molecular contour. Instead, a part of the torsional strain will be absorbed in the form of writhe~\cite{full71}, which, simply put, is a form of chiral bending. As long as the total torsional strain is small and the extending force sufficiently large~\cite{mark95b,moro98,mark15}, writhe manifests in small helical fluctuations around the fully extended state. Beyond a certain force-dependent threshold torsional strain, these fluctuations compound to initiate a buckling transition, which leads part of the molecule to wrap around itself in superhelical conformations called plectonemes~\cite{mark95b,mark12,eman13,skor22}.  In this section, we are only interested in the former regime.

Absorption of torsional strain into writhe results in an effectively reduced torsional stiffness—usually referred to as effective torsional stiffness $\ceff$, relative to the angular fluctuations of the molecular termini~\cite{nomi17},
\begin{equation}
    \label{eq:measure_ceff}
    \left\langle \Delta \theta^2 \right\rangle = \frac{L}{\ceff},
\end{equation}
where $\theta$ is the total angle traced out by the magnetic bead which is equal to the torsional fluctuation of the molecular endpoint. At low forces, writhe fluctuations are appreciable, leading to values of  $\ceff$ that are significantly lower than DNA's innate torsional stiffness $C$. Conversely, at large forces, bending fluctuations are largely suppressed, such that rotations of the magnetic bead are predominantly dictated by the torsional stiffness of the molecule. Moroz and Nelson showed that in the large force regime (once again $f \gtrapprox 0.3$~pN for literature values of DNA elastic parameters) the effective torsional stiffness can be expanded as~\cite{moro97,moro98}
\begin{equation}
    \label{eq:ceff_mn}
    \ceff = C - \frac{C^2}{4\lb}\sqrt{\frac{\kt}{\lb f}} + \ho
\end{equation}
Theoretical treatment in the low-force regime is also possible~\cite{bouc98,bouc00}, albeit not in closed form. However, such analysis is beyond the scope of this work. 

\subsubsection{Limitations arising from external forces}
The coarse-graining procedure eliminates original junctions via integration under the assumptions of absence of external constraints, i.e., the free energy of the original and coarse-grained system are equivalent only if all deformations in the corresponding segment are thermally activated and unrestrained. A molecule subject to a linear external force does not satisfy these requirements as the force effectively biases junction fluctuations. In principle, one would have to include this bias when carrying out the integration. However, the degree to which linear forces modify thermal fluctuations strongly depends on the wavelength of the fluctuating mode. Coupling to large wavelength fluctuations is strongest, while short wavelength modes remain almost unrestrained. For semi-flexible polymers the high-force correlation length of tangential bending fluctuations is~\cite{mark95,mark15,schi21}
\begin{equation}
    \label{eq:high_force_correlation_length}
    \xi = \sqrt{\frac{\kt A }{ f}}.
\end{equation}
This essentially sets the length scale beyond which the force suppresses bending fluctuations. Conversely, for fluctuations at given length scales, in particular the size of the composite segments, \Eqref{eq:high_force_correlation_length} provides an estimate of the threshold force beyond which fluctuations within the segment are appreciably attenuated.

\begin{figure*}[ht!]
\centering
\includegraphics[width=17.6cm]{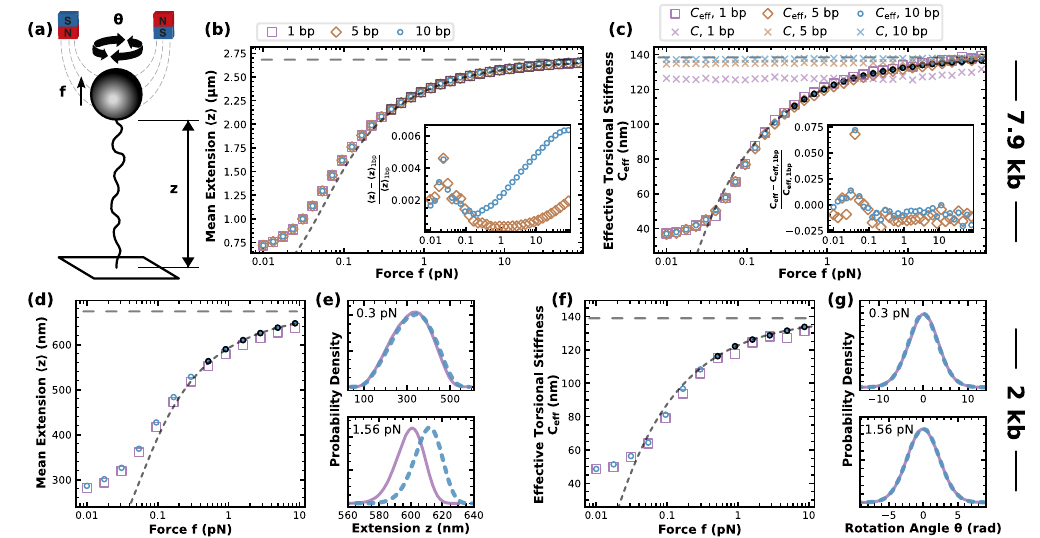}
\caption{Monte Carlo simulations of freely orbiting magnetic tweezer setup. (a): An illustration of a freely orbiting magnetic tweezer. (b,c): Comparison of force-extension (b) and the effective torsional stiffness (c) for simulations of $7.9$~kb molecules for three different resolutions. The underlying energy for these simulations only contains the rotational degrees of freedom tilt, roll, and twist and keeps the discretization length constant. Relative differences between the coarse-grained results from the single base pair-step resolution simulations are highlighted in the insets. 
The curved dashed line in panel (b) represents the interpolation formula, \Eqref{eq:forceext_interp}, obtained by fitting \Eqref{eq:forceext_highforce} to the 1~bp resolution data for forces exceeding $0.3$~pN.
The horizontal dashed line indicates the length of the molecule, marking the maximum tether extension.  In panel (c), the curved dashed line represents a fit of \Eqref{eq:ceff_mn} to the $10$~bp resolution data again for forces larger than $0.3$~pN. Here, the horizontal dashed line indicates the fitted value of $C$, which is asymptotically approached by $\ceff$ at large forces. The x-scatter points represent direct torsional stiffness $C$ measurements, calculated from the total twist fluctuations according to \Eqref{eq:measure_C}. 
(d-g): $2$~kb simulations including the translational degrees of freedom shift, slide, and rise.
Data corresponding to the $1$~bp resolution are shown in purple and those corresponding to $10$~bp resolution are shown in blue. Panels (e) and (g) show probability distributions of extension and rotation angle $\theta$, respectively, for two forces.
}
\label{fig:fomt} 
\end{figure*}

\subsubsection{Monte Carlo Simulations}
We simulated the FOMT setup for the same $7.9$~kb sequence~\cite{lipf10,krie18,vand22} used for the calculation of persistence lengths using the Markov Chain Monte Carlo method previously employed in the study of DNA plectonemes~\cite{vand22,skor22} (for more information see also Supplement of ~\cite{vand22}; code available at \href{https://github.com/eskoruppa/PolyMC}{https://github.com/eskoruppa/PolyMC}). The underlying energetic model is identical to the RBP description discussed here, with the limitation that it only considers rotational fluctuations and assumes a constant discretization length and no lateral translational components at the single junction level. 
We supplement the crankshaft and modified-pivot moves used in these works with a pivot move that allows for rotations of the terminal triad around its tangent. The orientation of said tangent remains constant and aligned with the force director field throughout the simulation, such that tracing the lateral triad vectors ($\uvec{u}$ and $\uvec{v}$ from \Eqref{eq:def_triads}) gives direct access to the global torsional angle $\theta$. Note that this procedure requires strand-crossing moves to be rejected since they would allow for the relaxation of torsional strain by other means than the rotation of the terminal triad. Moreover, the magnetic bead and the flow cell surface are explicitly considered in the form of two impenetrable surfaces anchored at the termini, to avoid linking number changes via crossings beyond said termini. 

We considered three separate resolutions, the original base pair-step parameterization, and coarse-grained representations at the $5$- and $10$-bp level ($k=5$ and $k=10$). 
Coarse-graining to 5~bp and 10~bp resolutions resulted in approximately 4-fold and 6-fold speedups of the MC simulation, respectively. Additionally, the sampling efficiency of the lower-resolution simulations is further enhanced by enabling larger MC moves.

\subsubsection{Force extension curves for $7.9$~kb MC simulations}
Force extension curves for these three respective resolutions and forces ranging from $0.01$~pN to $83$~pN are shown in \Figref{fig:fomt}(b). Base pair-step resolution extensions are faithfully reproduced by the coarse-grained simulations to the point that extension curves are visually indistinguishable. Closer inspection reveals relative differences to remain below $1\%$ (see inset of \Figref{fig:fomt}(b)). 
Relative differences exhibit an increasing tendency for large forces past a certain resolution-dependent threshold. This observation is in line with the discussion about the force-mediated suppression of lateral fluctuations as wavelengths below the respective discretization lengths commence to be attenuated by the stretching force. 
As expected, the threshold force is the lowest for the $10$~bp resolution simulations. Calculation of the apparent persistence length via fitting of \Eqref{eq:forceext_highforce} for forces larger than $0.3$~pN yields $\lb=68.4\pm0.2$~nm for the original resolution simulation. This value is in excellent agreement with the asymptotic persistence length calculated via the tangent-tangent correlation function. Fitting of the coarse-grained data yields slightly lower values: $\lb=67.7 \pm 0.3$~nm for the $5$~bp resolution and $\lb=65.3 \pm 0.4$~nm for the $10$~bp resolution. Interestingly, imposing an upper limit on the force range used for the fit recovers the base pair resolution persistence length. We find $\lb=68.3 \pm 0.1$~pN at $10$~bp resolution when only considering forces lower than $1$~pN and $\lb=68.4 \pm 0.2$~pN at $5$~bp resolution when only considering forces lower than $30$~pN. Again these findings are in line with the permeation of force-induced suppression of lateral fluctuations below the length scale of the coarse-graining resolution. The dashed line in \Figref{fig:fomt}(b) is a plot of the interpolation formula \Eqref{eq:forceext_interp} using the value of the persistence length deduced from the high force fit (for the $1$~bp resolution). Clearly, \Eqref{eq:forceext_interp} is not a good representation of the observed extension behavior for low forces as it predicts convergence to zero extension, while the simulations converge to a finite value. The low force behavior of \Eqref{eq:forceext_interp} is valid for ideal flexible polymers, i.e., for non-self-avoiding polymers, and in the absence of the repulsion planes. In particular, the presence of the latter breaks the symmetry of the distribution at zero force thereby necessitating a positive  and distinctly  non-zero mean extension.

\subsubsection{Effective torsional stiffness for $7.9$~kb MC simulations}
Values for $\ceff$ calculated from the same simulations are displayed in \Figref{fig:fomt}(c). Once again, the agreement between base pair-resolution and coarse-grained simulations is excellent with relative differences not exceeding $2.5\%$ (except for a single point; see inset of \Figref{fig:fomt}(c)). Fitting of the $1$~bp data for forces larger than $0.3$~pN to \Eqref{eq:ceff_mn} yields $C = 139.8 \pm 0.3$~pN and $\lb = 67.2 \pm 0.8$~pN. This value of $C$ may also be viewed as the high-force asymptote of $\ceff$ is indicated by the horizontal dashed line in \Figref{fig:fomt}(c). Quite remarkably, it agrees well with the direct measurements via the asymptotic long length-scale torsional persistence length for the coarse-grained representations calculated in the previous section (especially the $42$~bp and $21$~bp resolutions) but much less well with the results of the base pair-resolution model. The FOMT simulations give direct access to the twist fluctuations, which can then be related to the torsional stiffness via
\begin{equation}
    \label{eq:measure_C}
    \langle \Delta\tw^2 \rangle = \frac{L}{4 \pi^2 C}.
\end{equation}
The resulting values of $C$ (x-scatters in \Figref{fig:fomt}(c)) are independent of the force, as expected since there is no direct coupling between force and twist. Moreover, direct calculations of $C$ stemming from the coarse-grained simulations closely align with the previously indicated asymptote (horizontal dashed line), while the base pair-resolution data underestimates the torsional stiffness. We conclude that the appropriate value for the torsional stiffness (for the given RBP parametrization) is about $140$~nm.

Fitting of the coarse-grained simulations to \Eqref{eq:ceff_mn} yields almost the same values as the $1$~bp resolution simulations. At $10$~bp resolution we find $C = 138.4 \pm 0.4$~pN, and $\lb = 66.5 \pm 0.3$~pN, while the $5$~bp resolution simulations yield $C = 138.2 \pm 0.3$~pN, and $\lb = 66.0 \pm 0.7$~pN. Reducing the force range used for the fit slightly increases the obtained estimates for $\lb$.

\subsubsection{Inclusion of translational degrees of freedom - $2$~kb MC simulations}
For the inclusion of the translational degrees of freedom in the simulations, we used a Python implementation of the MC package (available at \href{https://github.com/eskoruppa/PMCpy}{https://github.com/eskoruppa/PMCpy}). However, its current limitations in computational efficiency made the generation of appreciable statistics for the $7.9$~bp sequence unattainable. Therefore we limited the simulations to a $2$~kb sequence taken as the first $2$~kb fragment of the $7.9$~kb sequence. Force-extension curves (\Figref{fig:fomt}(d)) and values of $\ceff$ (\Figref{fig:fomt}(f)) remain in reasonably agreement across $1$~bp and $10$~bp resolution simulations. Closer inspection reveals systematic discrepancies of the extension at large forces, where the distribution of $z$-values is markedly shifted to the upside for the coarse-grained representation (see (\Figref{fig:fomt}(e)). This observation aligns with the prior observed pathology arising from the coarse-graining of the rise. The coarse-graining procedure systematically overestimates composite rise components by not accounting for the emergent asymmetry of composite-step rise distributions (see \Figref{fig:local_composites}(a), and (c)). Distributions of the bead angle $\theta$, however, are almost perfectly reproduced by the coarse-grained representations (\Figref{fig:fomt}(g)).


\subsection{Superhelically curved DNA}
\label{sec:helix}

\begin{figure*}[t]
\centering
\includegraphics[width=17.6cm]{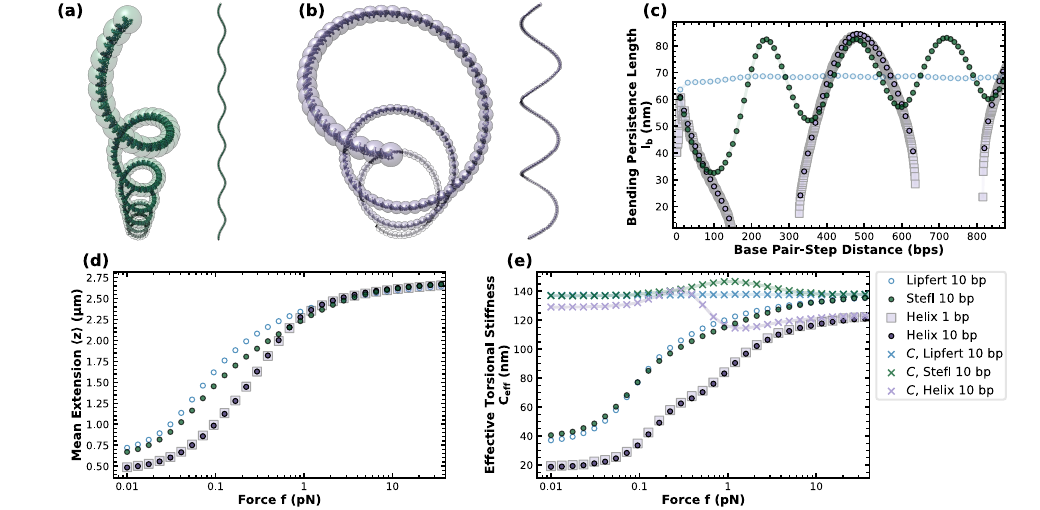}
\caption{
Simulations with superhelically curved sequences. Panels (a) and (b) depict the top and side view of the ground state of the phased A-tract sequences CAAAATTTTG~\cite{Stefl2004} and the phased sequence CGGGGGCTTTTAGGGGGCTTTTAGGGGGCTTT, respectively. (c): Comparison of the length scale-dependent bending persistence lengths for the two superhelical sequences and the previously considered experimental sequence~\cite{lipf10}. Force-extension and effective torsional stiffness data sampled with the FOMT setup for the same sequences are shown in (d) and (e), respectively. 
}
\label{fig:helix} 
\end{figure*}

As an example of a system where intrinsic curvature is manifestly relevant, we consider DNA sequences with an intrinsic helically wound ground state (see Figs~\ref{fig:helix}(a), and (b)). Such superhelical 
sequences can be constructed by repeatedly concatenating small segments of strong intrinsic curvature. We consider two such sequences. The first consists of the phased A-tracts CAAAATTTTG and was considered by Stefl et al.~\cite{Stefl2004}. Its ground state is shown in \Figref{fig:helix}(a). The second sequence, consisting of repeating segments of CGGGGGCTTTTAGGGGGCTTTTAGGGGGCTTT was constructed in the present work by searching for sequences with a prominent helical diameter, see \Figref{fig:helix}(b). 
For both superhelical sequences we constructed $7.9$~kb long molecules and conducted MC simulations in the unrestrained ensemble and the FOMT setup.

Figure~\ref{fig:helix}(c) shows length scale-dependent persistence lengths as given by \Eqref{eq:lb} for the two superhelical sequences and the previously considered $7.9$~kb sequence (taken from Lipfert et al.~\cite{lipf10}). Unsurprisingly, the large-scale helical structure of the two superhelical sequences manifests in strong sinusoidal oscillations of the apparent persistence length. Missing points in the curve correspond to base pair-step distances $m$ for which the tangent-tangent correlation function assumes negative values such that the logarithm in \Eqref{eq:lb} is undefined. 

In \Figref{fig:helix}(d) the force-extension curves of the two superhelical sequences are compared with the largely featureless $7.9$~kb sequence~\cite{lipf10}. Both sequences exhibit markedly different elastic responses. Specifically, the artificially constructed sequence deviates from the response of the reference sequence for forces lower than approximately 1~pN. In this force range, the elastic behavior is influenced not only by entropy but also by the Hookean response of the superhelix, see also Ref.~\cite{tomp16}. At higher forces, the superhelix is predominantly stretched out, leading the response to revert to that of the WLC model. 
The effective torsional stiffness $\ceff$ is even more distinct for the strongly superhelical sequence, exhibiting a lower stiffness for all forces. This behavior persists to the largest considered forces.


\section{Conclusions}
\label{sec:discussion}

In this work, we presented a systematic coarse-graining scheme that enables the calculation of sequence-specific coarse-grained parameters. These parameters faithfully reproduce structural and elastic properties relevant to the chosen resolution. The original system is assumed to be described as a chain of rigid bodies, each captured by an intrinsic reference frame and characterized by a ground state and a stiffness matrix. Importantly, our procedure does not predict elastic parameters but transforms a given set of RBP  parameters~\cite{olso98,lank03,lank09_rigidbp,petk14,sharma23} to a lower resolution.

To achieve a $k$-fold reduction in resolution, we retained every $k$-th base pair associated reference frame while eliminating all other frames. The ground state structure is then described in terms of the average relative orientation and position of these remaining reference frames. Fluctuations were captured through an effective stiffness matrix constructed to closely mimic the relative fluctuations of corresponding frames in the original system. Coarse-graining, therefore, entails mapping parameters from one Gaussian system into another. An extension of the coarse-graining procedure to incorporate higher-order effects, such as, for example, elastic multimodalities~\cite{dans12,liebl21,lopez23}, kinks~\cite{lank06,mitc11,volog13}, and linear sub-elastic behavior~\cite{wigg06,voorspoels2023}, is possible in principle but requires more sophisticated approaches that go beyond the scope of the current work.

A Python module featuring implementations for all parameter transformations discussed in this work is available at https://github.com/eskoruppa/PolyCG. These transformations included conversions between Cayley- and Euler-map definitions of rotations (see \appref{app:cayley}), as well as between midstep-triad definitions of translations and the definition relative to the base pair triad employed in this work (see \appref{app:midstep}). 

Benchmark simulations of unrestrained molecules based on RBP parameters derived from the cgNA+ model~\cite{sharma23} demonstrated that the coarse-grained parameters excellently capture the distributions of rotational degrees of freedom.  The variances of coarse-grained rotational degrees of freedom were found to deviate from numerically obtained references by less than 2\% for up to 40-fold reductions in resolution. 
However, translational fluctuations, particularly in rise, are less accurately reproduced. These discrepancies were attributed to the emergent asymmetry of composite-step translational fluctuations. Specifically, distributions of the total rise within a segment comprising multiple base pair steps exhibit marked left skewness, indicating that the end-to-end distance is more frequently contracted than extended. We argue this asymmetry to be a consequence of bending fluctuations, which, due to the entropic elasticity of a polymer~\cite{doi86}, results in end-point contractions. Moreover, since DNA is found to be nearly inextensible for forces below about $10$~pN~\cite{bust94}, it is reasonable to assume rotational fluctuations to constitute the most relevant component of sequence-dependent elasticity under physiological conditions. 

Sequence-specific and length scale-dependent signatures in the bending persistence length were shown to be faithfully reproduced by the coarse-grained systems. Furthermore, we demonstrated that the coarse-graining approach makes it possible to simulate setups typically studied with single-molecule techniques, where large molecules containing thousands of base pairs are probed. We showed simulations of freely orbiting magnetic tweezers at $10$~bp resolution to be virtually identical to equivalent simulations performed at full single-base pair resolution.
This agreement is retained even if sequences with strong intrinsic curvature—in this case, a superhelically would groundstate—are considered.
The coarse-graining approach, therefore, gives access to the sequence-dependent study of far beyond previously attainable length scales. 

The proposed methodology is not limited to a particular model or parameter set and can, in principle, be applied to achieve coarse-grained representations for any base pair resolution (or higher) model. Switching to coarser representations may enable multi-resolution simulations, capturing the region of interest in full detail while simulating the rest of the system at a lower resolution. Moreover, this approach could be used to expedite the equilibration of large systems.

We envision the access to faithful coarse-grained representations of DNA, provided by our coarse-graining approach to enable the in silico study of many sequence-specific mesoscale phenomena. 
In particular, coarse-grained simulations could aid in elucidating the role of DNA sequences on the statistics and dynamics of DNA plectonemes. An implementation of the coarse-grained RBP model in a Molecular Dynamics framework might further enable the study of chromosomal organization guided by DNA protein interactions such as restriction factors~\cite{vand22}, nucleosomes~\cite{Neipel2020,brac14_nuc}, and the localization of DNA loops~\cite{batt24}, for example, guided by the action of Structural Maintenance of Chromosomes (SMC) complexes~\cite{orla19,nomi22}.






\section{Acknowledgments}
Insightful discussions with Midas Segers and Stefanos K. Nomidis are gratefully acknowledged. ES and HS were supported by the Deutsche Forschungsgemeinschaft (DFG, German
Research Foundation) under Germany's Excellence Strategy - EXC-2068 -
390729961.


\appendix


\section{Rodrigues' Rotation Formula}
\label{app:rodrigues}

Matrix exponentials as in \Eqref{eq:expmap} are computationally expensive and in practice it is more convenient to use alternative formulations, such as Rodrigues' rotation formula
\begin{equation}
    \label{eq:rodrigues}
    \rot =  \eul{\vOm} := \ident + (\sin\Omega ) \hOm + (1-\cos\Omega)\hOm^2.
\end{equation}
The inverse transformation, which maps a rotation matrix to the corresponding rotation vector, is
\begin{equation}
    \label{eq:inv_rodrigues}
    \vOm = \ieul{\rot} := \frac{\Omega\, \vecmap\left(\rot - \rottp\right)}{2\sin \Omega},
\end{equation}
where $\Omega$ is determined by the relation $\tr \rot = 1 + 2\cos \Omega$.


\section{Conversion between Cayley map and Euler map definition of rotations}
\label{app:cayley}

Various studies~\cite{lank09_rigidbp,gonz13,petk14,debruin18,sharma23} favor a definition of rotational components based on the Cayley map (also known as the Euler–Rodrigues or Gibbs formula), 
\begin{equation}
    \label{eq:cayley_map}
    \rot = \cay{\vec{\Theta}} := \ident + \frac{4}{4+\vec{\Theta}^2}\left( \hm{\Theta} + \frac{1}{2}\hm{\Theta}^2\right),
\end{equation}
where we denote the Cayley vectors by $\vTh$ to distinguish them from the Euler-vectors $\vOm$. The inverse relation takes the simple form
\begin{equation}
    \label{eq:inv_cayley_map}
    \vTh = \icay{\rot} = \frac{2\,\vecmap\left(\rot - \rottp\right)}{1+\tr \rot} .
\end{equation}
Computationally these definitions have an edge over the rotation vector definition chosen in the main text (in that its definition does not involve any infinite series expansions---or transcendental functions), but in most practical situations the difference is rather inconsequential.

The vector $\vTh$ still indicates the axis of rotation, however, the rotation angle is no longer simply given by the vector's magnitude, but instead $|\vec{\Omega}| = 2\arctan(\Theta/2)$. This implies that the transformation of Cayley vectors into Euler vectors is given by
\begin{equation}
    \label{eq:cay2euler}
    \vec{\Omega} = \cayeul(\vec{\Theta}) := 2\arctan\left(\frac{\Theta}{2}\right) 
    \frac{\vec{\Theta}}{|\vec{\Theta}|},
\end{equation}
and the corresponding inverse by
\begin{equation}
    \label{euler2cay}
    \vec{\Theta} = \eulcay(\vec{\Omega}) := 2\tan\left(\frac{\Omega}{2}\right) 
    \frac{\vec{\Omega}}{|\vec{\Omega}|}.
\end{equation}
Suppose that the geometry and Gaussian elasticity of a given molecule containing $N+1$ base pairs is expressed in terms of a Cayley parametrization. The ground state will be specified by the system (Cayley-) vector 
\begin{equation}
    \tThs^\tp = 
    \begin{pmatrix}
        \vThs[,1]^\tp & \dots & \vThs[,N]^\tp
    \end{pmatrix},
\end{equation}
where $\vThs[,i]$ is the intrinsic rotation between base pairs $i$ and $i+1$ expressed in Cayley coordinates. Analogous to \Eqref{eq:X_energy_total}, the (lowest order expansion of the) elastic energy takes a (quadratic) form
\begin{equation}
    \label{eq:E_cay}
    \beta E = \frac{1}{2} \tThd^\tp M_{\Theta} \tThd.
\end{equation}

Transformation of the static component from Cayley to Euler representation is straightforward
\begin{equation}
    \vOms = \cayeul(\vThs).
\end{equation}
To identify how the stiffness matrix $M_{\Theta}$ transforms under the coordinate transformation, we expand $\cayeul$ to linear order around the ground state
\begin{equation}
    \label{eq:eulcay_linearexp}
    \Omega_\mu = \cayeul[,\mu](\vThs) + F_{\mu\nu} \Theta_{\Delta,\nu} + O(\Theta^2),
\end{equation}
where $\mu$ and $\nu$ indicate the dimensional subscripts of the single junction element, with summation over repeating indices implied. The positional subscript was omitted for ease of readability. Equations~\eqref{eq:eulcay_linearexp} implies that up to linear order
\begin{equation}
    \Omega_{\Delta,\mu} = F_{\mu\nu} \Theta_{\Delta,\nu},
\end{equation}
with
\begin{align}
    F_{\mu\nu} 
    &= \left.\frac{\partial \cayeul[,\mu] (\vec{\Theta})}{\partial \Theta_\nu}\right|_{\vec{\Theta}=\vThs} \n
    &= \left( \frac{4}{|\vThs|^{2} + 4} - \frac{|\vOms|}{|\vThs|} \right) \frac{\Theta_{0,\mu} \Theta_{0,\nu}}{|\vThs|^2} + \frac{|\vOms|}{|\vThs|} \delta_{\mu\nu},
\end{align}
where $\delta_{\mu\nu}$ is the Kronecker delta and $|\vOms|$ is given by \Eqref{eq:cay2euler}. Defining the block-diagonal matrix
\begin{equation}
    \label{euler_cayley_totaltrans}
    \totaltrans{F} = 
    \begin{pmatrix}
        F^{(1)} & 0 & \hdots & 0 \\
        0 & F^{(2)} & \hdots & 0 \\
        \vdots & \vdots & \ddots & \vdots \\
        0 & 0 & \hdots & F^{(N)} \\
    \end{pmatrix},
\end{equation}
that captures the transformation of the entire system, the transformation from system-wide Cayley to system-wide Euler vector may be written as 
\begin{equation}
    \tOmd = \totaltrans{F} \, \tThd.
\end{equation}
The elastic energy, \Eqref{eq:E_cay}, may then be rewritten as
\begin{align}
    \beta E 
    &=  \frac{1}{2} \tThd^\tp M_{\Theta} \tThd \n
    &= \frac{1}{2} \tOmd^\tp \left[ \left(\totaltrans{F}^{-1}\right)^\tp M_{\Theta} \totaltrans{F}^{-1} \right] \tOmd,
\end{align}
where the inverse of $\totaltrans{F}$ is also block-diagonal with entries explicitly given by
\begin{equation}
    F_{\mu\nu}^{-1} 
    = \left( \sec^2\left( \frac{|\vOms|}{2}\right) - \frac{|\vThs|}{|\vOms|} \right) \frac{\Omega_{0,\mu} \Omega_{0,\nu}}{|\vOms|^2} + \frac{|\vThs|}{|\vOms|} \delta_{\mu\nu}.
\end{equation}
Finally, we can identify the sought transformation of the stiffness matrix $M_\Theta$,
\begin{equation}
    \label{eq:cayley2euler_transform_stiffmat}
    M_{\Omega} = \left(\totaltrans{F}^{-1}\right)^\tp M_{\Theta} \totaltrans{F}^{-1}.
\end{equation}

In the present discussion, we assumed the energy to be composed of solely rotational components, but since the transformation from Cayley to Euler parameters leaves translations unaltered, the extension of the transformation of the stiffness matrix to include translations is straight-forward (note however, that the entries of the stiffness matrix quantifying translational stiffness do not generally remain unaltered).

Lastly, we note that the Jacobian associated with the transformation, $\cayeul$,
\begin{equation}
    \det G = \frac{4 |\vOms|^2}{(|\vThs|^2+4)|\vThs|^2},
\end{equation}
takes values markedly different from $1$ for typical values of $\vThs$ (for double-stranded DNA) but varies by less than $5\%$ over typical ranges of $\vTh$, justifying it to be ignored in most practical cases~\cite{thesis_davia}. 

\begin{figure}[t!]
\centering
\includegraphics[width=8.6cm]{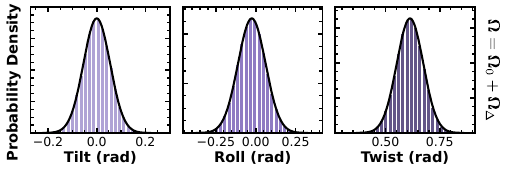}
\caption{Comparison between sampled and analytical distributions of Euler-vector component for a single base pair-step. The stiffness matrix in Cayley-representation was generated with cgNA+~\cite{sharma23} and corresponds to a single GpC step, which was generated in the context of a larger sequence (GACTGC\textbf{GC}GCCTCA) to avoid boundary effects. Tilt, Roll, and Twist were sampled in the original (Cayley) coordinates and then transformed to Euler coordinates via \Eqref{eq:cay2euler}.}
\label{fig:cayley2euler} 
\end{figure}

We validate the quality of the analytical transformation of the stiffness matrix by comparing numerically sampled distributions (generated according to the canonical measure), with distributions obtained by marginalizing the analytically transformed stiffness matrix. Values are sampled in the Cayley representation and then transformed to Euler-vectors via \Eqref{eq:cay2euler}. Comparative histograms for a single base pair-step revealing a relative difference in variance of less than $1\%$ are shown in \Figref{fig:cayley2euler}.


\section{Splitting static and dynamic components at the transformation level}
\label{app:algebra2group}

As introduced in the main text, it is customary to express the transformation between consecutive base pair-related frames as a single six-vector, containing three rotational and three translational components. In the compact SE(3) notation such a transformation takes the form
\begin{equation}
    g = 
    \begin{pmatrix}
        \rot & \vec{v} \\ 
        \vec{0}^\tp & 1 \\
    \end{pmatrix}
    =
    \begin{pmatrix}
        e^{\hOms + \hOmd} & \vvs + \vvd \\ 
        \vec{0}^\tp & 1 \\
    \end{pmatrix}.
\end{equation}
However, the coarse-graining procedure discussed in the main text requires static and dynamic components to be split at the transformation level, i.e.,
\begin{equation}
    g = s\, d,
\end{equation}
with 
\begin{equation}
    \label{eq:app_def_s}
    s = 
    \begin{pmatrix}
        \srot & \vs \\ 
        \vec{0}^\tp & 1 \\
    \end{pmatrix}
    =
    \begin{pmatrix}
        e^{\hPhs } & \vs \\ 
        \vec{0}^\tp & 1 \\
    \end{pmatrix},
\end{equation}
and 
\begin{equation}
    \label{eq:app_def_d}
    d = 
    \begin{pmatrix}
        \drot & \vd \\ 
        \vec{0}^\tp & 1 \\
    \end{pmatrix}
    =
    \begin{pmatrix}
        e^{\hPhd } & \vd \\ 
        \vec{0}^\tp & 1 \\
    \end{pmatrix}.
\end{equation}
In this section, we will show how to achieve this decomposition and how the stiffness matrix transforms under such redefinition of fluctuating variables. 

For consistency, the ground state should be identical across both pictures, which leads to $\vs = \vvs$ and $\vPhs = \vOms$. Therefore, we can treat $s$ as a known quantity, such that $d$ may be expressed as
\begin{equation}
    \label{eq:expr_d}
    d = s^{-1} g =
    \begin{pmatrix}
        \srot^\tp \rot & \srot^\tp \left(\vec{v} - \vec{s}\right) \\ 
        \vec{0}^\tp & 1 \\
    \end{pmatrix}
    =
    \begin{pmatrix}
        \srot^\tp \rot & \srot^\tp \vec{v}_\Delta \\ 
        \vec{0}^\tp & 1 \\
    \end{pmatrix}.
\end{equation}
From \Eqref{eq:expr_d} we can deduce that
\begin{equation}
    \label{eq:relation_vdel_and_d}
    \vec{d} = \srot^\tp \vvd,
\end{equation}
and 
\begin{equation}
    \label{eq:relation_hPhd_hOms_hOmd}
    e^{\hPhd} = e^{-\hOms} e^{\hOms + \hOmd}.
\end{equation}
The exact correspondence between $\hOms$ and $\hPhd$ is encapsulated in the Baker-Campbell-Hausdorff formula, which for three elements $a$, $b$, and $c$ of a Lie algebra provides the solution of the equation $e^{a+b}=e^{c}$ in terms of the infinite series of commutators
\begin{align}
    \label{bch}
    c = 
    &\phantom{+}a + b + \frac{1}{2}[a,b] \n
    &+ \frac{1}{12}\left([a,[a,b]] + [b,[b,a]]\right) + \ho,
\end{align}
where h.o. is a placeholder for the infinite series of higher-order commutators. We seek the transformation from $\vOmd$ to $\vPhd$ up to linear order. Therefore, it will suffice to consider only terms linear in the fluctuating quantities. The first few terms in the Baker-Campell-Haussdorf formula satisfying this requirement are
\begin{align}
    \label{split_bch}
    \hOmd 
    &= \log \left( e^{\hOms} e^{\hPhd}\right) - \hOms \n
    &= \hPhd + \frac{1}{2} [\hOms,\hPhd] + \frac{1}{12}[\hOms,[\hOms,\hPhd]] \n
    &\phantom{=} -\frac{1}{720} [\hOms,[\hOms,[\hOms,[\hOms,\hPhd]]]] \n
    &\phantom{=} +\frac{1}{30240} [\hOms,[\hOms,[\hOms,[\hOms,[\hOms,[\hOms,\hPhd]]]] \n
    &\phantom{=} + \mathrm{h.o.}
\end{align}
Note that
\begin{equation}
    \vecmap\left([\hOms,\hPhd]\right) = \vec{\Omega}_0 \times \vPhd = \hOms\vPhd,
\end{equation}
nested commutators containing $k$ static elements $\hOms$ and only a single $\hPhd$, which is arranged to be the right-hand element in the innermost commutator, take a simple form when expressed as vectors instead of antisymmetric tensors
\begin{equation}
    \label{eq:nested_commutators}
    \vecmap\left([\hOms,[\hOms,[\ldots,[\hOms,\hPhd]\ldots]]]\right) = \hOms^k\vPhd.
\end{equation}
After substitution of \Eqref{eq:nested_commutators} into \Eqref{split_bch}, one has up to linear order
\begin{equation}
    \label{eq:phi_om_transformation}
    \vec{\Omega}_\Delta = H^{-1}(\vOms)\; \vPhd,
\end{equation}
with 
\begin{equation}
    \label{eq:def_Hinv}
    H^{-1} (\vOms)\; = \ident + \frac{1}{2}\hOms + \frac{1}{12}\hOms^2-\frac{1}{720}\hOms^4+\frac{1}{30240}\hOms^6 + \ho
\end{equation}
We note, that the Jacobian of this transformation is $1$ since $\det \hOms = 0$. 

Defining the six-component basepair-step deformation vectors that include both the rotational and the translational components 
\begin{equation}
    \vXd^\tp \equiv
    \begin{pmatrix}
        \vOmd^\tp & \vvd^\tp 
    \end{pmatrix} 
    \qquad
    \vYd^\tp \equiv
    \begin{pmatrix}
        \vPhd^\tp & \vd^\tp 
    \end{pmatrix} 
\end{equation}
one sees that the transformation from $\vXd$ to $\vYd$ may be linearized as
\begin{equation}
    \vYd = H\left( \vXs \right) \vXd,
\end{equation} 
where the just derived linearized transformations of rotation (\Eqref{eq:def_Hinv}) and translation (\Eqref{eq:relation_vdel_and_d}) are the respective components of the matrix
\begin{equation}
    H\left( \vXs \right) = 
    \begin{pmatrix}
        H(\vOms) & 0 \\
        0 & \srottp
    \end{pmatrix}.
\end{equation}
Defining 
\begin{equation}
    \bar{H}\left( \tXs \right) = 
    \begin{pmatrix}
        H\left( \vXs[,0] \right) & 0 & \hdots & 0 \\
        0 & H\left( \vXs[,1] \right) & \hdots & 0 \\
        \vdots & \vdots & \ddots & \vdots \\
        0 & 0 & \hdots & H\left( \vXs[,N-1] \right)
    \end{pmatrix},
\end{equation}
and following the same logic that led to \Eqref{eq:cayley2euler_transform_stiffmat}, one finds 
\begin{equation}
    \label{eq:algebra2group_transform_stiffmat}
    M_Y = \left[\bar{H}^{-1}\left( \tXs \right)\right]^\tp M_X \left[\bar{H}^{-1}\left( \tXs \right)\right].
\end{equation}


\section{Conversion between midstep-triad and triad definition of translations}
\label{app:midstep}

To achieve a parametrization of DNA configurations that are independent of the strand direction—up to a sign-flip—(i.e., assignment of Watson and Crick strand), translational base pair-step components are frequently expressed in terms of mid-step coordinate frames that are found via half-way rotation between the two respective reference frames (rotational components are identical when defined relative to either of the two respective triads or the mid-step frame). However, for this work, it proved convenient to express translations in the frame of the first triad of each respective pair of frames. In this section, we show how these definitions can be transformed into one another and how the associated transformation of the stiffness matrix can be accommodated.

The mid-step triad is usually defined as~\cite{lave09_curves} 

\begin{equation}
    \triad_{\text{mid},i} = \triad_i \sqrt{\rot_i},
\end{equation}
where $\sqrt{\rot_i}$ is the matrix of half rotation between the frames $\triad_i$ and $\triad_{i+1}$. Specifically, if 
\begin{equation}
    \rot_i = \triadtp_i \triad_{i+1} = \exp\hOm[i],
\end{equation}
then 
\begin{equation}
    \sqrt{\rot_i} \equiv \exp\frac{1}{2}\hOm[i].
\end{equation}
With respect to the mid-step triad $\triad_{\text{mid},i}$ the translation between base pairs $i$ and $i+1$, is given by~\cite{thesis_davia}
\begin{equation}
    \label{eq:def_zeta}
    \boldsymbol{\zeta}_i = \triadtp_{\text{mid},i} \left(\vec{r}_{i+1} - \vec{r}_i\right).
\end{equation}
Meanwhile, the triad frame definition as introduced in the main text is 
\begin{equation}
    \label{eq:def_v}
    \vec{v}_i = \triadtp_i \left(\vec{r}_{i+1} - \vec{r}_i\right).
\end{equation}
Inspection of Eqs.~\ref{eq:def_zeta} and \ref{eq:def_v} reveals the relation between the two definitions
\begin{equation}
    \label{eq:zeta2v}
    \vec{v}_i = \sqrt{\rot_i} \boldsymbol{\zeta}_i.
\end{equation}

To ascertain the transformation of the stiffness matrix, we will once again follow the procedure of first splitting the transformation into static and dynamic components, where the dynamic component is assumed to be small, followed by the linearization of the dynamic part of the transformation.

Following previous treatment the rotation $\sqrt{\rot_i}$ may be expanded to first order in the fluctuating rotational component $\vOmd[,i]$,
\begin{align}
    \sqrt{\rot_i} 
    &= \exp\left(\frac{1}{2}\hOms[,i] + \frac{1}{2}\hOmd[,i]\right) 
    \n
    &\approx \sqrt{\srot_i} \exp\left(\frac{1}{2} \hatmap \left[ H\left(\frac{\vOms[,i]}{2} \right) \vOmd[,i] \right] \right)
    \n
    &\approx \sqrt{\srot_i} \left( \ident + \frac{1}{2} \hatmap \left[ H\left(\frac{\vOms[,i]}{2} \right) \vOmd[,i] \right] \right),
\end{align}
where we used \Eqref{eq:phi_om_transformation} to split the exponential.
Separating $\boldsymbol{\zeta} = \boldsymbol{\zeta}_0 + \boldsymbol{\zeta}_\Delta$ into static and dynamic components \Eqref{eq:zeta2v} takes the form
\begin{equation}
    \vec{v}_i \approx \sqrt{\srot_i} 
    \left( 
    \boldsymbol{\zeta}_{0,i} + \boldsymbol{\zeta}_{\Delta,i} 
    + \frac{1}{2} \hatmap \left[ H\left(\frac{\vOms[,i]}{2} \right) \vOmd[,i] \right] \boldsymbol{\zeta}_{0,i}
    \right).
\end{equation}
Following previous procedure the quadratic term coupling $\boldsymbol{\zeta}_{\Delta,i}$ and $\vOmd[,i]$ was ignored. Finally, using 
\begin{equation}
    \label{eq:switch_antisymm_vec}
    \hm{a}\vec{b} 
    = \vecmap \left( \left[\hm{a}, \hm{b} \right]\right)
    = \vecmap \left( \left[{\hm{b}}^\tp, \hm{a} \right]\right) 
    = \hm{b}^\tp \vec{a},
\end{equation}
one arrives at 
\begin{align}
    \label{eq:transform_midstep2triad}
    \vvs[,i] &= \sqrt{\srot_i} \boldsymbol{\zeta}_{0,i},
    \\
    \vvd[,i] &= 
    \sqrt{\srot_i}\boldsymbol{\zeta}_{\Delta,i} 
    + 
    \frac{1}{2} \sqrt{\srot_i} \hm{\zeta}_{0,i}^\tp H\left(\frac{\vOms[,i]}{2} \right) \vOmd[,i].
\end{align}
The fluctuating component $\vvd[,i]$ does not solely depend on the $\boldsymbol{\zeta}_{\Delta,i}$, but rotational fluctuations $\vOmd[,i]$ lead to additional translational fluctuation. In matrix form, the transformation takes the form
\begin{equation}
    \label{eq:midstep_transformation}
    \begin{pmatrix}
        \vOmd[,i] \\ 
        \vvd[,i]
    \end{pmatrix}
     = 
    \begin{pmatrix}
        \ident & 0 \\ 
        \frac{1}{2} \sqrt{\srot_i} \hm{\zeta}_{0,i}^\tp H\left(\frac{\vOms[,i]}{2} \right) & \sqrt{\srot_i}
    \end{pmatrix}
    \begin{pmatrix}
        \vOmd[,i] \\ 
        \boldsymbol{\zeta}_{\Delta,i}
    \end{pmatrix}.
\end{equation}
The transformation of the stiffness matrix is then completely analogous to \Eqref{eq:algebra2group_transform_stiffmat}.

\begin{figure}[t!]
\centering
\includegraphics[width=8.6cm]{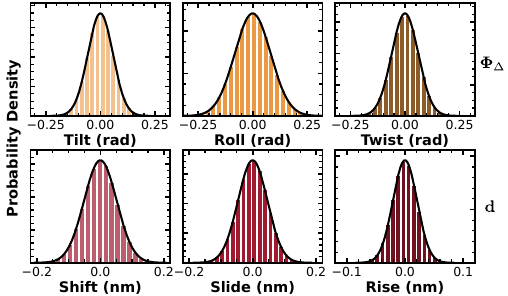}
\caption{Comparison between sampled and analytical distributions of the dynamic components of tilt, roll, and twist (in the $\vPhd$ parametrization; expressed in radians) and shift, slide, and rise (in the $\vd$ parametrization; expressed in nm) for a single base pair-step. The same sequence and procedure were considered as in \Figref{fig:cayley2euler}.}
\label{fig:algebra2group} 
\end{figure}

We jointly assess the quality of the transformations of \appref{app:algebra2group} and \appref{app:midstep} in the same way as was done for the transformation of the stiffness matrix from the Cayley map definition of rotations to the Euler map definition (see \appref{app:cayley}). Rotations and translations were sampled according to the $\vOmd$ and $\boldsymbol{\zeta}_{\Delta}$ parametrizations respectively and then transformed to $\vPhd$ and $\vd$ via Eqs.~\eqref{eq:zeta2v}, \eqref{eq:relation_vdel_and_d}, and \eqref{eq:relation_hPhd_hOms_hOmd}. Comparisons of the resulting histograms with the analytical predictions obtained via Eqs.~\eqref{eq:midstep_transformation} and \eqref{eq:algebra2group_transform_stiffmat} followed by the marginalization of the resulting stiffness matrix to the effective stiffness of the respective degrees of freedom are shown in \Figref{fig:algebra2group}. Deviations between sampled and analytical variances were not found to exceed $2\%$ for any considered sequence out of a total sample of 500 sequences.


\section{Composite Translations}
\label{app:comp_trans}

This section serves to provide the details of the calculation of the composite translation (\Eqref{eq:comp_trans}). 
Working out the translational component of
\begin{equation}
    \dse_{\comp{i}{j}} 
    = \sseinv_{\accu{i}{j}} \gseph_{\accu{i}{j}},
\end{equation}
in terms of single junctions rotations and translations one finds

\begin{align}
    \label{eq:dij_bare}
    \vd[\comp{i}{j}] 
    &= \srottp_{\accu{i}{j}} \vv[\accu{i}{j}] - \srottp_{\accu{i}{j}} \sum_{l=i}^{j} \srotph_{\accu{i}{l-1}} \vs[l]
    \n
    &=
    \srottp_{\accu{i}{j}} \sum_{l=i}^{j} \left[ \rot_{\accu{i}{l-1}} \left( \srot_l \vd[l] + \vs[l]  \right) \right] - \sum_{l=i}^{j} \srottp_{\accu{l}{j}} \vs[l].
\end{align}
Applying the approximation used for the calculation of the rotational component (see Section~\ref{sec:ad_rot}) then yields

\begin{align}
    \label{eq:apptrans_expand_STR}
    \srottp_{\accu{i}{j}} \rotph_{\accu{i}{l-1}} 
    &= 
    \srottp_{\accu{l}{j}} \left( \srottp_{\accu{i}{l-1}} \rotph_{\accu{i}{l-1}} \right)
    \n
    &= 
    \srottp_{\accu{l}{j}} \drotph_{\comp{i}{l-1}} 
    \n
    &\approx
    \srottp_{\accu{l}{j}} \exp \left( \hatmap \left(  \sum_{k=i}^{l-1} \srottp_{\accu{k+1}{l-1}} \vPhd[,k]  \right) \right)
    \n
    &\approx
    \srottp_{\accu{l}{j}} \left( \ident +  \hatmap \left(  \sum_{k=i}^{l-1} \srottp_{\accu{k+1}{l-1}} \vPhd[,k]  \right) \right)
\end{align}
where the matrix exponential was expanded to first order in the last step. Substituting \Eqref{eq:apptrans_expand_STR} into \Eqref{eq:dij_bare} one further finds
\begin{widetext}
\begin{align}
    \label{eq:dij_2_terms}
    \vd[\comp{i}{j}] 
    &\approx
    \sum_{l=i}^j 
    \left[
    \srottp_{\accu{l}{j}} \left( \ident +  \hatmap \left(  \sum_{k=i}^{l-1} \srottp_{\accu{k+1}{l-1}} \vPhd[,k]  \right) \right)
    \left( \srot_l \vd[l] + \vs[l]  \right) 
    \right]
    - \sum_{l=i}^{j} \srottp_{\accu{l}{j}} \vs[l]
    \n
    &\approx
    \sum_{l=i}^j  \srottp_{\accu{l+1}{j}} \vd[l] 
    + 
    \sum_{l=i}^j \srottp_{\accu{l}{j}} \hatmap \left(  \sum_{k=i}^{l-1} \srottp_{\accu{k+1}{l-1}} \vPhd[,k]  \right) \vs[l] 
    \n
    &= 
    \sum_{l=i}^j  \srottp_{\accu{l+1}{j}} \vd[l] 
    + 
    \sum_{l=i}^j \sum_{k=i}^{l-1} \srottp_{\accu{l}{j}} \hatmap \left( \srottp_{\accu{k+1}{l-1}} \vPhd[,k]  \right) \vs[l].
\end{align}
From the first to the second line, the terms coupling $\vPhd[k]$ to $\vd[l]$ were discarded since they are of quadratic order in the fluctuations. Moreover, the constant term, $\sum_{l=i}^{j} \srottp_{\accu{l}{j}} \vs[l]$, exactly cancels out.

Equation~\eqref{eq:dij_2_terms} shows the dynamic composite translation, $\vd[\comp{i}{j}]$, to depend on both the dynamics junction translation, $\vd[l]$, and the dynamic junction rotation, $\vPhd[,l]$. However, to identify the proper junction rotation transformation one has to rewrite the second term in \Eqref{eq:dij_2_terms} as 
\begin{align}
    \label{eq:switch_hatmaps}
    \srottp_{\accu{l}{j}} \hatmap \left( \srottp_{\accu{k+1}{l-1}} \vPhd[,k]  \right) \vs[l] 
    &= \srottp_{\accu{l}{j}} \hat{s}_{l}^\tp \srottp_{\accu{k+1}{l-1}} \vPhd[,k]
    \n
    &= \srottp_{\accu{l}{j}} \hat{s}_{l}^\tp \srotph_{\accu{l}{j}} \srottp_{\accu{k+1}{j}} \vPhd[,k]
    \n
    &= \hatmap \left (\srottp_{\accu{l}{j}} \vs[l]^\tp\right) \srottp_{\accu{k+1}{j}} \vPhd[,k],
\end{align}
where \Eqref{eq:switch_antisymm_vec} was used in the first step to shift $\vPhd[,k]$ to the left-hand side of the expression. After combining Eqs.~\eqref{eq:dij_2_terms} and \eqref{eq:switch_hatmaps} and rearranging the order of the summation one finally arrives at 
\begin{equation}
    \label{eq:trans_final}
    \vd[\comp{i}{j}] = \sum_{l=i}^j  \srottp_{\accu{l+1}{j}} \vd[l]  + \sum_{k=i}^{j-1} \left[ \sum_{l=k+1}^j  \hatmap\left( \srottp_{\accu{l}{j}} \vs[l]^\tp \right) \srottp_{\accu{k+1}{j}} \right] \vPhd[,k].
\end{equation}
\end{widetext}



%

\end{document}